%% file: main.tex
\documentclass[11pt]{article}

\usepackage[american]{babel}
\usepackage{natbib}
\usepackage[figuresleft]{rotating}
\usepackage{mathtools} 
\usepackage{booktabs} 
\usepackage{tikz} 
\usepackage{adjustbox}
\usepackage{multirow}
\usepackage{amsmath}
\usepackage{amssymb}
\usepackage{mathtools}
\usepackage{amsthm}
\usepackage{mathabx} 
\usepackage{bbm}
\usepackage{microtype}
\usepackage{graphicx}
\usepackage{subcaption}
\usepackage{booktabs} 
\usepackage[colorlinks=true, allcolors=blue]{hyperref}
\usepackage[capitalize,noabbrev]{cleveref}
\RequirePackage{algorithm}
\RequirePackage{algorithmic}
\RequirePackage{forloop}
\RequirePackage{url}

\usetikzlibrary{shapes.geometric}

\theoremstyle{plain}
\newtheorem{theorem}{Theorem}[section]

\newtheorem{fact}[theorem]{Fact}

\newtheorem{prop}{Proposition}

\theoremstyle{definition}

\newtheorem{assumption}[theorem]{Assumption}
\theoremstyle{remark}

\usepackage[textsize=tiny]{todonotes}
\newcommand{\ind}{\perp\!\!\!\perp}

\title{Scalable Causal Structure Learning via Amortized Conditional Independence Testing}
\author{
James Leiner\textsuperscript{1} , Brian Manzo\textsuperscript{3}, Aaditya Ramdas\textsuperscript{1,2} , Wesley Tansey\textsuperscript{4} 
}
\date{}

\begin{document}
\maketitle

\begin{tabular}{l}
\textsuperscript{1}Department of Statistics and Data Science, Carnegie Mellon University \\
\textsuperscript{2}Machine Learning Department, Carnegie Mellon University \\
\textsuperscript{3}Department of Statistics, University of Michigan \\
\textsuperscript{4}Computational Oncology, Memorial Sloan Kettering Cancer Center \\
\end{tabular}

\begin{center}
\today
\end{center}
\begin{abstract}
\input{CLEAR_2025/abstract}
\end{abstract}
\input{CLEAR_2025/intro}
\input{CLEAR_2025/method}
\input{CLEAR_2025/results}
\input{CLEAR_2025/conclusion}
\bibliography{references}
\appendix

\input{CLEAR_2025/appendix}

\end{document}

%% file: CLEAR_2025/abstract.tex


Controlling false positives (Type I errors) through statistical hypothesis testing is a foundation of modern scientific data analysis. Existing causal structure discovery algorithms either do not provide Type I error control or cannot scale to the size of modern scientific datasets. We consider a variant of the causal discovery problem with two sets of nodes, where the only edges of interest form a bipartite causal subgraph between the sets.
We develop Scalable Causal Structure Learning (SCSL), a method for causal structure discovery on bipartite subgraphs that provides Type I error control.
SCSL recasts the discovery problem as a simultaneous hypothesis testing problem and uses discrete optimization over the set of possible confounders to obtain an upper bound on the test statistic for each edge.
Semi-synthetic simulations demonstrate that SCSL scales to handle graphs with hundreds of nodes while maintaining error control and good power.
We demonstrate the practical applicability of the method by applying it to a cancer dataset to reveal connections between somatic gene mutations and metastases to different tissues.

%% file: CLEAR_2025/intro.tex
\section{Introduction} \label{sec:intro}

Many scientific applications can be posed as a causal discovery problem where a directed acylic graph (DAG) is learned that models the causal dependencies within an observational dataset (e.g. \cite{application1}, \cite{application2}). In order to ensure the reliability of these findings, controlling the error rate on the set of causal discoveries is critical. Given finite data and noisy observations, exact determination of causal arrows in a DAG is impossible. As such, the causal discovery task is typically cast through the lens of statistical hypothesis testing for conditional independencies~\citep{spirtes:etal:2000:pc-algorithm}.

Although the problem of large-scale multiple hypothesis testing and uncertainty quantification is well studied in settings where purely associative relationships are the target discoveries \citep{bh_procedure,goeman_hyptesting}, large-scale causal learning produces a unique set of challenges. The number of possible DAGs scales superexponentially with the number of nodes in a graph, making causal learning an especially difficult problem when the number of variables in a dataset is large. Moreover, existing causal structure learning algorithms learn a graph with either no attempt to provide frequentist error guarantees \citep{GES_orig, Ramsey2017AMV,zheng2018dags,cundy2021bcd,Annadani2021VariationalCN,BCD_NETS} or require parametric assumptions on the data generating distribution which are unknown in practice \citep{strobl:etal:2016:pc-p-algorithm}. 

A feature of many scientific datasets that can be exploited to decrease the computational burden of this task is \emph{temporal separation} of variables. If a dataset has a sequence of variables that measure quantities that came into existence at different times, this corresponds to a priori knowledge that some edges can only be oriented in a particular direction. This allows us to reduce the number of conditional independence tests required to draw an edge on a causal graph. 

In this paper, we introduce Scalable Causal Structure Learning (SCSL), a method for large-scale causal hypothesis testing that can scale to problems with hundreds of variables and thousands of potential edges for causal graphs with temporally separated sets of nodes. SCSL enables the use of black box machine learning models for hypothesis testing, requires no parametric assumptions, and returns a $p$-value for each edge under consideration. To scale to large graphs, SCSL recasts the causal search process as a discrete optimization problem. It then amortizes the dominant cost in causal structure identification: conditional independence testing over the combinatorial set of possible parent nodes in the causal graph. This avoids the combinatorial explosion in the candidate conditioning sets and reduces the search to a series of parallelizable optimization problems for each edge.
We validate SCSL through semi-synthetic experiments using a cancer dataset that pairs genomic mutations at the primary tumor location of a cancer patient with information about metastases that have developed elsewhere in a patient's body \citep{data_source}. These simulation studies indicate that the method has high power, controls Type I error rate at the target level, and scales to larger graphs than existing methods.~\looseness=-1

\paragraph{Background and related work} Classical algorithms for causal discovery like the SGS and PC algorithms (\cite{spirtes:etal:2000:pc-algorithm}) convert causal structure learning into queries of conditional independence between nodes. The SGS algorithm draws an edge between two nodes if they are conditionally dependent given any possible subset of the remaining nodes. The PC algorithm simplifies this process by first deleting edges in the causal graph based on marginal independence testing. The size of the conditioning subsets is then allowed to increase by one for each subsequent round of CI testing, but fewer CI tests are required in each round as the algorithm only considers conditioning subsets of nodes that have remained adjacent to each other. The computational complexity of SGS matches the complexity of the PC in the worst case, but in most settings where the causal graph is reasonably sparse, the PC algorithm significantly reduces the number of independence tests needed to learn a causal graph. However, both methods are based on perfect knowledge of the CI structure and are computationally intractable beyond a few dozen nodes. Related work used the PC algorithm as a starting point and then relaxes assumptions around faithfulness (e.g. \cite{ramsey_conservative_pc}) or causal sufficiency (e.g. \cite{FCI}), but these methods also rely on having perfect knowledge of the graph's CI structure. The work of \cite{strobl:etal:2016:pc-p-algorithm} extends the PC algorithm to generate edge-specific $p$-values, but only with provable Type I error control under the assumption of zero Type II error during the edge deletion stage of skeleton discovery.

Other approaches to causal graph are score-based methods which  maximize a score function such as BIC \citep{BIC} or BDeau \citep{bdeau} over the space of all possible DAGs. Since the number of DAGs increases superexponentially with the number of nodes, approximate algorithms based on greedy search \citep{GES_orig,Ramsey2017AMV}, coordinate descent \citep{coord_aragam,coord_qing} or other methods are required. Other methods impose more stringent modeling assumptions such as linearity (e.g. \cite{LINGAMdirect}) simplify the task. Follow-up work \citep{fges,BOSS} improves several of these methods to scale to datasets with thousands of nodes and potential edges using parallelization. However, none of these methods output edge specific $p$-values and are therefore not directly comparable to our proposed method. 


More recently, differentiable approaches to causal discovery have been proposed \citep{zheng2018dags}, many of which enable Bayesian inference of the posterior distribution over DAGs \citep{cundy2021bcd,Annadani2021VariationalCN,BCD_NETS}. Unfortunately, these methods focus primarily on continuous rather than discrete data, limiting the application to datasets like the cancer dataset we consider in our simulations. Further, Bayesian uncertainty requires accurate model specification for valid posterior coverage. When the model is misspecified, Bayesian credible intervals can massively inflate Type I error. Similarly, \citet{icp_algo} weaken the faithfulness assumptions, but require repeated observations of variables across different intervention settings instead of relying on purely observational data. Overall, no method exists that is capable of providing Type I error control on individual edges across a large graph without making stringent parametric assumptions.

%% file: CLEAR_2025/method.tex
\section{Methodology}\label{sec:problem}
Consider a directed acyclic graph (DAG) $\mathcal{G}$ with two sets of nodes, $\mathcal{X}$ and $\mathcal{Y}$, where we want to learn the directed edges between $\mathcal{X}$ and $\mathcal{Y}$ from observational data. We assume that we have prior knowledge that no edge is directed from $\mathcal{Y}$ to $\mathcal{X}$ such as temporal separation between the sets of nodes. For instance, in the cancer dataset, $\mathcal{Y}$ would represent metastasis events that occur after the tumors represented by $\mathcal{X}$ were sequenced.


The problem reduces to learning which $X_j \in \mathcal{X}$ are connected to which $Y_k \in \mathcal{Y}$. A naive approach might be to infer an edge based on a conditional independence test of $X_{j} \ind Y_{k} | \{ X_{-j}, Y_{-k}\}$ where $X_{-j} := \mathcal{X}\setminus X_{j}$ and $Y_{-k} := \mathcal{Y} \setminus Y_{k}$. However, a collider being present in $\mathcal{Y}$ could introduce dependence even when there is no edge present as \Cref{fig:abxy-graph} demonstrates.

\begin{figure}[ht]
    \centering
    \begin{tikzpicture}[
roundnode/.style={circle, draw=black!60, very thick, minimum size=5mm},
]
\node[roundnode]      (lowerleftcircle)                              {$X_1$};
\node[roundnode]        (upperleftcircle)       [above=of lowerleftcircle] {$X_2$};
\node[roundnode]      (rightbottomcircle)       [right=of lowerleftcircle] {$Y_1$};
\node[roundnode]        (righttopcircle)       [right=of upperleftcircle] {$Y_2$};

\draw[thick, ->] (upperleftcircle) -- (lowerleftcircle);
\draw[thick, ->] (upperleftcircle) -- (righttopcircle);
\draw[thick, ->] (righttopcircle) -- (rightbottomcircle);
\draw[thick, ->] (lowerleftcircle) -- (rightbottomcircle);
\end{tikzpicture}
    \caption{$\mathcal{X} = \{X_{1}, X_{2} \}$ and $\mathcal{Y} = \{Y_{1}, Y_{2} \}$. $Y_{1}$ is a collider, so a conditional independence test would indicate $X_{1}$ and $ Y_{2}$ are conditionally dependent given $\{X_{2}, Y_{1}\} $. However, $X_{1}$ and $Y_{2}$ are not conditionally independent given  $X_{2}$. Querying all dependence relations prevents an edge from being drawn erroneously.}
    \label{fig:abxy-graph}
\end{figure}
We make the following assumptions about the graph and distribution of the data which are common in the literature \citep{spirtes:etal:2000:pc-algorithm}. 

\begin{assumption}[Global directed Markov property] \label{assumption:markov}
For a graph $\mathcal{G}$, if two nodes $U$ and $V$  are d-separated given a disjoint set nodes $W$, then $U$ and $V$ are conditionally independent given $W$. 
\end{assumption}

\begin{assumption}[d-separation Faithfulness] \label{assumption:dsep}
For a graph $\mathcal{G}$, if two nodes $U$ and $V$ are conditionally independent given a disjoint set of nodes $W$, then $U$ and $V$ are d-separated given $W$.
\end{assumption}

\begin{assumption}[Causal sufficiency]  \label{assumption:sufficiency}
The graph $\mathcal{G}$ includes all common causes for any pair of nodes contained in $\mathcal{G}$. 
\end{assumption}

These assumptions lead to \cref{prop:edge}, which enables us to reduce queries about edge presence to queries about conditional independence between nodes.

\begin{prop} \label{prop:edge}
Assume $\mathcal{G}$ satisfies the global directed Markov property and the probability distribution is d-separation faithful. Furthermore, assume that edges may not be directed from any element in $\mathcal{Y}$ to any element in $\mathcal{X}$. Then there is an edge between two vertices $X_{j} \in \mathcal{X}$ and $Y_{k} \in \mathcal{Y}$ if and only if $X_{j}$ and $Y_{k}$ are conditionally dependent given $S \cup X_{-j}$ for all $S \subseteq Y_{-k}$. \end{prop}

In order to construct a p-value, we consider a hypothesis test for each pair $(X_{j},Y_{k})$ as follows:
\begin{align} \label{eqn:hyp_edge}
\begin{split}
&H_{0}: X_{j} \rightarrow Y_{k} \text{ is absent,} \\
&H_{1}: X_{j} \rightarrow Y_{k} \text{ is present.} 
\end{split}
\end{align} 
\Cref{prop:edge} lets us restate the null and alternative as 
\begin{align} \label{eqn:hyp_edge2}
\begin{split}
&H_{0}:\exists S \subseteq \mathcal{Y} \setminus Y_{k} \text{ such that }  X_{j} \ind Y_{k} | S,  X_{-j}, \\
&H_{1}:  X_{j} \text{ is not independent of } Y_{k} \text{ given } \{S, X_{-j} \} \text{ for all } S \subseteq Y_{-k}.\\
\end{split}
\end{align} 
Let $p_{X_{j} \ind Y_{k} | S}$ denote the p-value corresponding to a test for conditional independence between $X_{j}$ and $Y_{k}$ given a set of nodes $S$.
\Cref{eqn:hyp_edge2} implies that the p-value $p_{X_{j} \rightarrow Y_{k}}$ can be bounded by
\begin{equation} \label{eqn:pvalue_bound}
p_{X_{j} \rightarrow Y_{k}} \le \max_{S \subseteq Y_{-k}} p_{X_{j} \ind Y_{k} | S, X_{-j}}.
\end{equation}
Constructing a $p$-value to test $H_{0}$ thus reduces to computing $p$-values for a series of conditional independence tests. 

\subsection{Conditional Independence Testing} \label{sec:GCM}


To test the null hypothesis that $X_{j} \ind Y_{k} | S, X_{-j}$, we employ the generalized covariance measure (GCM) of \cite{hardness_CI}. The GCM recasts the problem of conditional independence into one about functional estimation of the expected conditional covariance
\begin{align} \label{eqn:pvalue_optim}
\begin{split}
\mathbb{E} \left[\text{Cov}\left( X_{j}, Y_{k} |  S, X_{-j} \right)\right] := & \mathbb{E}[\mathbb{E}[X_{j}Y_{k}| S, X_{-j}] \\ 
- &\mathbb{E}[X_{j}| S, X_{-j}] \mathbb{E}[Y_{k}| S, X_{-j}] ],
\end{split}
\end{align}
and then tests whether this quantity is equal to $0$. Any set of joint densities for $(\mathcal{X}, \mathcal{Y})$ that are null  will have this property, although it is possible for $X_{j}$ and $Y_{k}$ to be conditionally dependent but with $0$ covariance. Therefore, this measure will only have power against the set of alternatives which have non-zero conditional covariance, which is likely to be the case in most practical settings. 

The procedure constructs a test for the null hypothesis of $0$ conditional covariance by first finding estimates $\hat{X}_{j}$ for the conditional expectation $\mathbb{E}[X_{j} | S, X_{-j}]$ and  $\hat{Y}_{k}$ for the conditional expectation $\mathbb{E}[Y_{k}| S, X_{-j}]$. Any type of predictive modeling to construct this estimate is valid, for example construction of a neural net, so long as the product of the mean square errors of the two quantities is $o(n^{-1})$. Given  $n$ samples $\{X_{j}^{i}, Y_{k}^{i}, S^{i}, X_{-j}^{i}\}_{i=1}^{n}$, we denote $R_{i} = \left( X_{j}^{i} - \widehat{X}_{j}^{i} \right) \left( Y_{k}^{i} - \widehat{Y}_{k}^{i} \right)$, and the test statistic
\begin{equation} \label{eqn:test_statistic}
    T^{(n)}=\frac{\sqrt{n} \cdot \frac{1}{n} \sum_{i=1}^n R_i}{\left(\frac{1}{n} \sum_{i=1}^n R_i^2-\left(\frac{1}{n} \sum_{r=1}^n R_r\right)^2\right)^{1 / 2}},
\end{equation}
which will be distributed asymptotically as $\mathcal{N}(0,1)$ under the null, allowing us to construct a valid $p$-value. For a precise account of the technical conditions for the theorem, see Theorem~6 of \cite{hardness_CI} which we recall in the Appendix for completeness.

We note that our method is adaptable to other CI testing procedures such as conditional randomization tests (e.g.  \cite{model_X},  \cite{tansey_HRT}, \cite{distilled_CRT}). We focus on the GCM in this work because it requires less stringent modeling assumptions on the conditional distribution $X|Z$ and is less computationally burdensome.

\subsection{Amortized Predictive Modeling} \label{sec:pred}
The GCM only requires the computation of a single test statistic per conditional independence query. While the computation of the individual test statistics is straightforward, this approach may still become computationally onerous when testing for the presence of an edge using \Cref{eqn:pvalue_bound} due to the fact that two separate predictive models need to be created for each of the $2^{|\mathcal{Y}|-1}$ CI tests corresponding to $H_{0}$ which will cause this methodology to scale poorly as  $|\mathcal{Y}|$ increases.

To make this process more computationally tractable, we will create a single predictive model for each element $Y_{k} \in \mathcal{Y}$ that has the flexibility to take in a choice of conditioning set $S \subseteq Y_{-k}$ as well as a choice of element $X_{j} \in \mathcal{X}$ and outputs an estimate of $\mathbb{E}[Y_{k} | X_{-j},S]$. We call this \emph{amortized predictive modeling} because instead of creating bespoke models for each $S$, we localize the cost of training into a single flexible model. Formally, we wish to train a function for each $Y_{k}$ denoted as $\pi_{k} : \mathbb{R}^{|\mathcal{X}|} \times \mathbb{R}^{|\mathcal{Y}|} \times \{0,1\}^{|\mathcal{Y}| -1} \times [|\mathcal{X}|] \rightarrow \mathbb{R}$ where $[|\mathcal{X}|] := \{1,...,|\mathcal{X}|\}$. The first two inputs into the function are the realized data points corresponding to $\mathcal{X}$ and $\mathcal{Y}$ while the second two inputs are user-specified masks which correspond to the  set of nodes that the user wishes to include in the conditioning set. For the third input, which we label $Y^{\text{mask}}$, we interpret $S := \{  Y_{i}^{\text{mask}} \text{ s.t. } Y^{\text{mask}}_{i} = 1 \}$. For the fourth input, we can interpret the choice as corresponding to an element $X_{j} \in \mathcal{X}$ to \emph{not} include in the conditioning set $X_{-j}$.  


When the predictive model is trained by minimizing a loss function $\ell$ through gradient descent, our strategy will be to amortize the model by randomly masking the inputs when using mini-batching. In particular, for each mini-batch, we can 
sample $B_{k} \in \text{Ber}(p)$ and $M \in \text{Cat}(|\mathcal{X}|,q)$ where the probabilities $q$ are just $\frac{1}{|\mathcal{X}|}$ for each component. Then, the sampled data $X^{i}_{j}$ and $Y^{i}_{k}$ are replaced with perturbed updates:
$$\widetilde{Y}^{i}_{k} := Y^{i}_{k} \times B_{k} , \text{ and } \widetilde{X}^{i}_{j} := X^{i}_{j} \times M_{i},$$
where  $M_{i} := \mathbbm{1}_{M \ne j}$. The parameter $p$ can be chosen by the user based on the amount of dependency they believe is present within $\mathcal{Y}$. For sparser graphs, the user may wish to choose larger values of $p$ because there is less potential for colliders so a larger conditioning set is more appropriate. If the user suspects $\mathcal{Y}$ to be less sparse, they can choose smaller values of $p$ to bias towards smaller conditioning sets. This process is summarized in \cref{alg:batch_grad} and can be visualized in \cref{fig:backprop}.

\begin{algorithm}[tb]
   \caption{Amortized predictive model training} \label{alg:batch_grad}
\begin{algorithmic}
   \STATE {\bfseries Input:} Data: $\mathcal{D}_{\mathcal{X}} \in \{-1,1\}^{n \times |\mathcal{X}|}$; $\mathcal{D}_{\mathcal{Y}} \in \{-1,1\}^{n \times |\mathcal{Y}|}$; $n_{ep}$ (number of epochs), $n_{batch}$ (batch size), $p$ (masking parameter), loss function $\ell \left(\theta, \mathcal{X}, Y_{-k}, Y_{k} \right)$
    \FOR{$i=1$ {\bfseries to} $n_{ep}$}
      \FOR{$i=1$ {\bfseries to} $[n/n_{batch}]$}
       \STATE Draw $n_{batch}$ new samples
       \STATE Draw $B_{m} \sim \text{Ber}(p)$ for each $Y_{m} \in Y_{-k}$
       \STATE Draw $M \sim\text{Cat}(\mathcal{X},q)$ with equally-weighted probabilities
       \STATE Construct $\tilde{Y}_{-k}$ by taking each $Y^{i}_{m}$ in the sample and replacing with $\widetilde{Y}^{i}_{k} := Y^{i}_{m} \times B_{m}$
       \STATE Construct $\tilde{\mathcal{X}}$ by taking each $X^{i}_{j}$ and replacing with $\widetilde{X}^{i}_{j} := X^{i}_{j} \times \mathbbm{1}_{M \ne k}$
       \STATE Compute $\frac{\partial \ell(\tilde{\mathcal{X}}, \tilde{Y}_{-k}, Y_{k})}{\partial \theta}$
       \STATE $\theta \leftarrow \theta - \eta  \frac{\partial \ell(\tilde{\mathcal{X}}, \tilde{Y}_{-k}, Y_{k})}{\partial \theta}$ 
    \ENDFOR
   \ENDFOR
\end{algorithmic}
\end{algorithm}

\tikzstyle{arrow} = [thick,->,>=stealth]
\tikzstyle{feat} = [circle, text centered, draw=black]

\tikzstyle{model} = [rectangle, text centered, draw=black,minimum height =1cm]
\tikzstyle{elli} = [text centered]

\begin{figure}[ht] 
    \centering
\begin{tikzpicture}
\node (x1) [feat] {$X_{1}$};
\node (elli) [elli,below = 0.1cm of x1] {\vdots};
\node (xp) [feat,below = 0.2cm of elli] {$X_{p}$};

\node (y1) [feat,below = 0.2cm of xp] {$Y_{1}$};
\node (elli2) [elli,below = 0.1cm of y1] {\vdots};
\node (yp) [feat,below = 0.2cm of elli2] {$Y_{m}$};

\node (m1) [feat,left = 0.6cm of x1] {$M_{1}$};
\node (elli3) [elli,below = 0.1cm of m1] {\vdots};
\node (mp) [feat,below = 0.2cm of elli3] {$M_{p}$};

\node (b1) [feat,left = 0.6cm of y1] {$B_{1}$};
\node (elli2) [elli,below = 0.1cm of b1] {\vdots};
\node (bp) [feat,below = 0.2cm of elli2] {$B_{m}$};

\node (b1) [elli,left = 0.05cm of y1] {$\times$};
\node (bp) [elli,left = 0.05cm of yp] {$\times$};

\node (b144) [elli,left = 0.05cm of x1] {$\times$};
\node (b44r) [elli,left = 0.05cm of xp] {$\times$};

\node (mod) [model,fill=brown!30, right = 1cm of xp] {ML Model};

\node (pr) [model,below = 0.5cm of mod,minimum height =0.5cm] {$\hat{Y}_{k}$};

\node (loss) [model,below = 0.5cm of pr,minimum height =0.5cm] {$\ell(\hat{Y}_{k}, Y_{k})$};

\node (annotate) [right= 1.2cm of pr,minimum height =0.8cm] {Backpropagation};
\draw [arrow] (x1) -- (mod);

\draw [arrow] (xp) -- (mod);
\draw [arrow] (y1) -- (mod);
\draw [arrow] (yp) -- (mod);

\draw [arrow] (mod) -- (pr);

\draw [arrow] (pr) -- (loss);

\draw [arrow,dotted] (loss) to [out=0,in=0] (mod);
\end{tikzpicture}

\caption{Illustration of masking procedure with mini-batch gradient descent. During model training, masks $B_{k} \sim \text{Ber}(p)$ and $M \in \text{Cat}(|\mathcal{X}|,q)$ are sampled during mini-batching to randomly hide nodes within $\mathcal{Y}$. This simulates the process of a user choosing a conditioning subset during model evaluation.}\label{fig:backprop}
\end{figure}
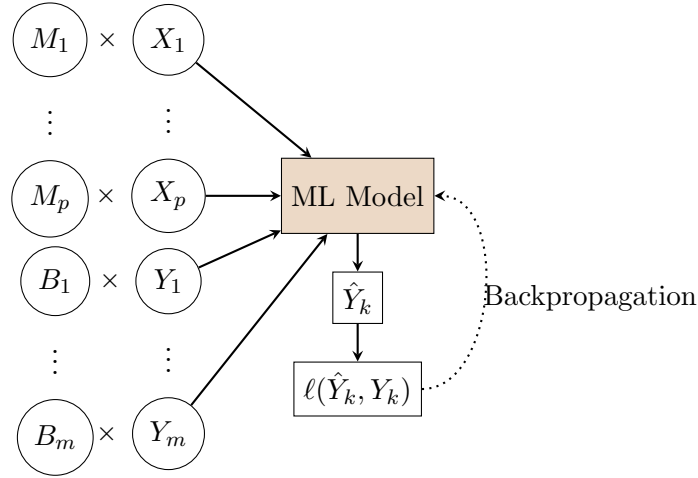

For simplicity, we assume all data is binary-valued. This allows us to only consider interactions between the generated masks and the elements of $\mathcal{X}$ and $\mathcal{Y}$. In the continuous setting, this can be generalized by including $\emph{both}$ the interactions and the masks themselves as inputs into the predictive model. Additional details and empirical results for the continuous case are included in the Appendix. 

Note that the above process describes a methodology for computing an estimate of $\mathbb{E}[Y_{k} | X_{-j},S ]$, but we need a similar procedure for computing an estimate of $\mathbb{E}[X_{j} | X_{-j}, S]$ for every $X_{j} \in \mathcal{X}$. We proceed in much the same way as before, but now let  $\pi_{j} : \{-1,1\}^{|\mathcal{X}|} \times \{-1,1\}^{|\mathcal{Y}|} \times \{0,1\}^{|\mathcal{Y}| -1}\rightarrow \mathbb{R}$. The user only needs to choose $S \subseteq \mathcal{Y}$ because $X_{-j}$ is conditioned on by default. \cref{alg:batch_grad} can then be modified to only draw Bernoulli variables and ignore the categorical variable used to mask the elements of $\mathcal{X}$. We again make this precise in the Appendix. 


\subsection{P-value optimization} \label{sec:pval_opt}

In order to find the maximal $p$-value corresponding to \Cref{eqn:pvalue_bound}, one approach would be to exhaustively search over all possible subsets. However, this will not be computationally tractable for large graphs as the number of conditional independence tests needed to compute a $p$-value for a single edge scales exponentially with the number of nodes. Building on recent work in differentiable casual structure learning \citep{brouillard2020differentiable,kalainathan2022structural,ng2022masked}, the approach we  pursue is to learn the conditioning subset that minimizes the test statistic in \Cref{eqn:test_statistic} through numerical optimization. Since the conditioning set is a discrete rather than continuous variable, standard techniques such as gradient descent do not immediately apply. To overcome this, we use the Gumbel-Softmax reparamaterization trick \citep{gumbel_softmax} to express the gradient with respect to the discrete variables we wish to optimize with continuous relaxations. 

Formally, we are learning the parameter $\theta := (\theta_1,...,\theta_m)$ where the conditioning subset is sampled as $\mathbbm{1}_{Y_{i} \in S} \sim \text{Ber}(\theta_{i})$. We then search for the value of $\theta$ which minimizes the expected value of \Cref{eqn:test_statistic} through gradient descent. To do this, we approximate $\frac{\partial T^{(n)}}{\partial S} \approx \frac{\partial T^{(n)}}{\partial \tilde{S}}  $ where $\tilde{S}$ is constructed as
$$\tilde{S}_{i} = \frac{\exp \left((\log \theta_i + g_{i1} \right)/\tau)  }{\exp \left(\log \theta_i + g_{i1} \right) + \exp \left(\log (1 - \theta_i) + g_{i2} \right) },$$
with $g_{i1}, g_{i2} \sim \text{Gumbel}(0,1)$. This approximates a discrete variable, with the quality of the approximation increasing as $\tau \rightarrow 0$. There's a trade-off between the quality of the approximation and the variance of the gradient, so we set $\tau$ to be large at first and then anneal it over subsequent iterations. The above process is summarized in \cref{fig:opt_procedure}. 

\tikzstyle{param} = [diamond, text centered, draw=black,fill=green!20]
\tikzstyle{rand} = [rectangle, rounded corners,text centered,draw=black,fill=blue!20,minimum height = 0.75cm,minimum width = 0.75cm]
\tikzstyle{ds2} = [rectangle,text centered, rounded corners, draw=black,minimum height = 0.75cm,minimum width = 1cm]
\tikzstyle{ds3} = [text centered, rounded corners]
\tikzstyle{ds} = [rectangle,text centered, rounded corners, minimum height=1cm,draw=black,minimum width = 3cm]
\tikzstyle{output} = [rectangle,text centered, rounded corners, minimum height=0.75cm,draw=black,minimum width = 3cm,fill=red!20]
\tikzstyle{output2} = [rectangle,text centered, rounded corners, draw=black,fill=red!20]
\tikzstyle{arrow} = [thick,->,>=stealth]

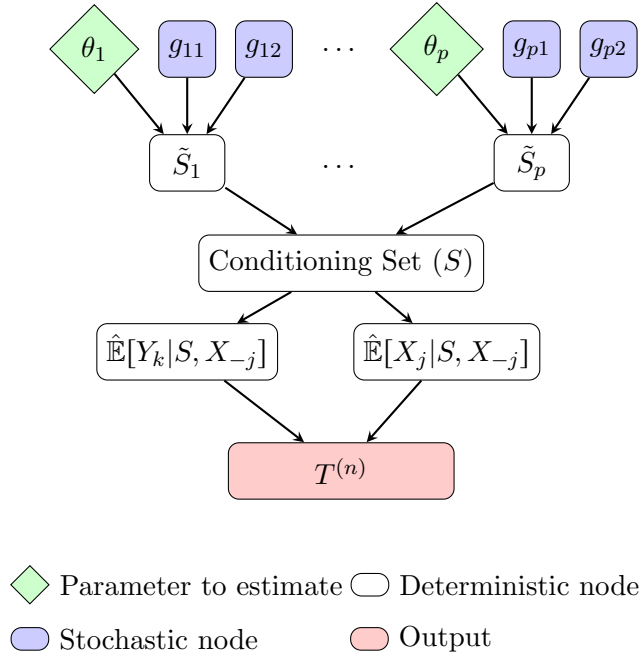
\begin{figure}[t!] \
    \centering
\begin{tikzpicture}
\node (theta) [param] {$\theta_{1}$};
\node (g11) [rand,right =0.25cm  of theta] {$g_{11}$};
\node (g12) [rand,right =0.25cm  of g11] {$g_{12}$};
\node (z) [ds2,below = 0.75cm of g11] {$\tilde{S}_{1}$};
\node (a) [ds3,right  =0.25cm of g12] { \dots };
\node (theta_p) [param,right =0.25cm  of a] {$\theta_{p}$};
\node (gp1) [rand,right =0.25cm  of theta_p] {$g_{p1}$};
\node (gp2) [rand,right =0.25cm  of gp1] {$g_{p2}$};
\node (z_p) [ds2,below = 0.75cm of gp1] {$\tilde{S}_{p}$};

\node (a1) [ds3,below  =1.25cm of a] { \dots };

\node (S) [ds2,below = 0.75cm of a1] { Conditioning Set ($S$)};
\node (predX) [ds2, below  = 1.75cm of z] {$\mathbb{\hat{E}}[Y_{k} | S, X_{-j}]$ };
\node (predY) [ds2, right = 1cm of predX] {$\mathbb{\hat{E}}[X_{j} | S, X_{-j}]$ };
\node (T) [output,  below = 2cm of S] {$T^{(n)}$ };

\node(legend_1) [param,below left  = 1cm  and 2.5cm of T,label=east:Parameter to estimate] {};
\node(legend_2) [rectangle,text centered,minimum width = 0.5cm,minimum height = 0.3cm, rounded corners, draw=black,right = 4cm of legend_1,label=east:Deterministic node] {};
\node(legend_3) [rand,below = 0.3cm of legend_1,minimum width = 0.5cm,minimum height = 0.3cm,label=east:Stochastic node] {};
\node(legend_4) [output2,right = 4cm of legend_3,minimum width = 0.5cm,minimum height = 0.3cm, label=east:Output] {};

\draw [arrow] (theta) -- (z);
\draw [arrow] (g12) -- (z);
\draw [arrow] (g11) -- (z);

\draw [arrow] (theta_p) -- (z_p);
\draw [arrow] (gp2) -- (z_p);
\draw [arrow] (gp1) -- (z_p);

\draw [arrow] (z) -- (S);
\draw [arrow] (z_p) -- (S);

\draw [arrow] (S) -- (predX);

\draw [arrow] (S) -- (predY);

\draw [arrow] (predX) -- (T);

\draw [arrow] (predY) -- (T);
\end{tikzpicture}
\caption{Optimization procedure. Stochastic nodes are sampled from a Gumbel distribution, allowing for back propagation of $\frac{\partial{T^{(n)}}}{\partial \theta_{i}}$.}
\label{fig:opt_procedure}
\end{figure}

As a final step, after some number of iterations where the parameter $\theta$ is learned, we need to define a procedure for converting the probabilities to a discrete set of choices $\hat{S}$. A successful search generates a $\hat{S}$ corresponding to a test statistic at least as small as the statistic for any d-separating set $S^{\star}$. We investigate two different approaches. First, we simply let  $\hat{S} = \{ i: \hat{\theta}_i > 0.5 \}$ after $q$ iterations. This is labeled \textbf{Gumbel-Softmax optimization (GSO)} in simulations. 

For the second approach, after $q_{1}$ iterations, we sample from conditioning subsets without replacement, but with the probability of sampling a set being proportional to its probability given $\hat{\theta}$. Precisely, the weights for each subset $S$ are chosen as $w_{S} = \prod_{i} \hat{\theta}_{i} ^{ \mathbbm{1}_{i \in S}}(1-\hat{\theta}_{i})^{ \mathbbm{1}_{i \not \in S}}$ for every $S \in \mathcal{P} (Y_{-k})$. After $q_{2}$ samples, $\hat{S}$ is taken to be the subset yielding the minimal test statistic over the $q_{2}$ samples. This is labeled \textbf{hybrid approach} in simulations. The hybrid approach tackles the search process in two stages. The first is an exploration stage where a probability distribution over the combinatorial space is learned. The second step explores this space efficiently by prioritizing sets that are most likely under the learned distribution. Empirically, this approach has better results than only using Gumbel-Softmax optimization or the more simplistic approach of sampling without replacement from all possible sets using equal weights, which we explore in detail in \cref{fig:performance_comp}.

In our experiments, we choose fixed values for the parameters $q$, $q_1$, and $q_2$ --- this corresponds to learning parameters with gradient descent and terminating the process after a fixed number of iterations. In principle, the performance of the method may be improved by using other stopping rules for gradient descent, such as terminating after the change in loss is below a threshold. We leave a detailed empirical investigation of this point an open avenue of inquiry. 

\paragraph{Early stopping rule} 
By definition, $p_{X_{j} \rightarrow Y_{k}}  \le 1$.
If during the search process, we find that there exists an $S$ such that $\hat{p}_{X_{j} \ind Y_{k} | S, X_{-j}} > \alpha$ for some pre-determined level $\alpha$ that we know we are not interested in rejecting above, we end the search early by setting $\hat{p}_{X_{j} \rightarrow Y_{k}} := 1$. This decreases the computational cost of computing $p$-value for null edges without impacting Type I error control and at the cost of power only at rejection thresholds that are of no practical significance. 

%% file: CLEAR_2025/results.tex
\section{Results} \label{sec:results}
We perform a benchmark and simulation study centered on a $n = 22,352$ dataset that pairs metastatic events with pre-metastasis tumor mutational info~(\cite{data_source}). Previous studies have looked at the genomic landscape of metastases in different tissues such as brain~\citep{brastianos2015genomic}) and breast \citep{brown2017phylogenetic} metastases. However, this dataset is the only large database that pairs metastatic events with pre-metastasis tumor mutational info across a range of different sites in the body. In total, $234$ genes were sequenced for each patient along with $23$ secondary metastatic tissue sites (e.g. colon, breast, brain, etc.). Although records are collated across dozens of primary tumor site locations, we focus on the $10$ sites with the most patient records availability (breast, colon, liver, lung, ovary, pancreas, rectum, skin, and uterus) to ensure adequate sample size.

This dataset presents an opportunity for statistical models to discover new genomic biomarkers of metastatic potential. 
For each sequenced gene in a given tumor, we wish to know whether a mutation in that gene has a causal effect on metastasis to another site. Somatic mutations in genes are often highly correlated~\citep{cheng2015memorial}, eliminating the use of simple association tests. Further, tumor colonization in one site may cause eventual metastasis to another site, such as liver metastases in colon cancer~\citep{paschos2009role}. It is therefore important to discover biomarkers with direct causal effect from the gene mutation to the specific metastatic site, such that intervention on that biomarker will have a positive effect on patient outcomes. Further, to ensure that discovered biomarkers are reliable, we wish to control the statistical error rate on reported causal links. 

Since the secondary tumor sites developed only after the primary tumor location has been sequenced, we know a priori that one set of variables (gene mutations in the primary tumor site) cannot be caused by a second set of variables (secondary metastatic events). We therefore can apply the methodology developed in this paper directly, letting $\mathcal{X}$ denote the mutations that have been sequenced and $\mathcal{Y}$ denote the potential secondary metastatic locations. 
\paragraph{Modelling approach} The underlying predictive models used to construct the GCM test statistics for all the results in this section are logistic regressions with $L_{2}$ regularization. Although we experimented with other approaches such as neural networks for learning the regression functions, we found these methods to have similar or lower power compared to a more parsimonious logistic model so these results were omitted. See the Appendix for further comparisons on test statistics constructed from alternative predictive models.
\paragraph{Baselines} Our primary baseline is the PC-p algorithm (\cite{strobl:etal:2016:pc-p-algorithm}) as it is the only other existing methodology designed for frequentist error coverage. We use the same GCM test statistics described to perform CI tests for this method. We also employ the same simplifying assumptions used by SCSL to streamline the number of conditional independence tests that need to be evaluated --- namely, by conditioning on all elements of $\mathcal{X}$ by default and orienting all edges between $\mathcal{X}$ and $\mathcal{Y}$ away from genetic sites and towards metastases. 

We also compare to other causal search algorithms implemented by the \textsc{TETRAD} project. Although these methods are not direct competitors to SCSL because they do not produce $p$-values and therefore cannot be used for false discovery rate control, empirical results suggests that SCSL's performance is still competitive. More information about the benchmark methodologies provided by \textsc{TETRAD} are included in the Appendix.

\begin{figure}
\centering
        \includegraphics[width=0.5\linewidth]{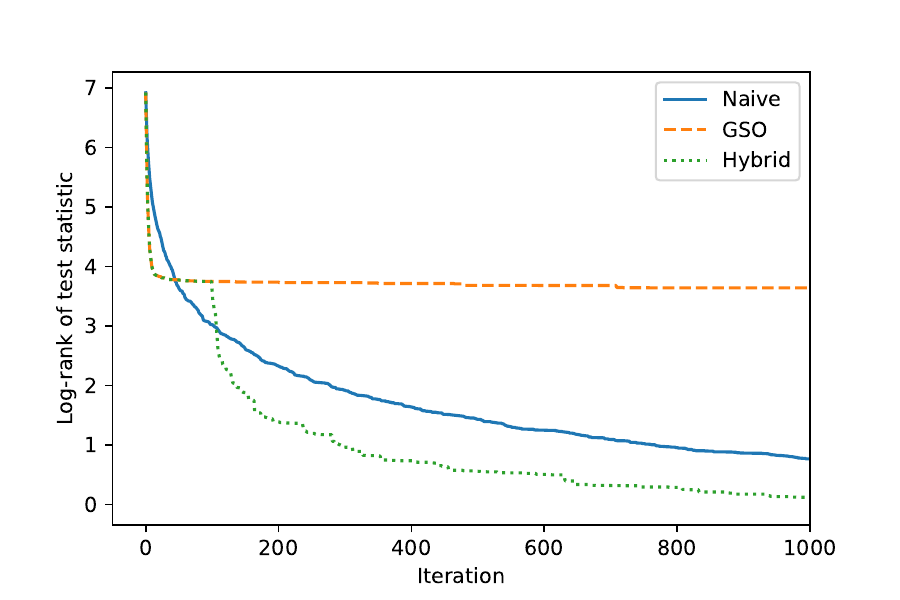}
    \caption{Comparison of the log-rank of the test statistic for each of the methodologies described in \Cref{sec:pval_opt} when run on a semi-synthetic dataset created as described in \cref{sec:semisynth_synthetic} with $|\mathcal{X}| = 47$ and $|\mathcal{Y}| = 12$. 
    The hybrid approach dominates both the naive and GSO approaches after a sufficient number of iterations.}
    \label{fig:performance_comp}
\end{figure}

\begin{figure*}
\centering 
      \includegraphics[width=0.44\linewidth]{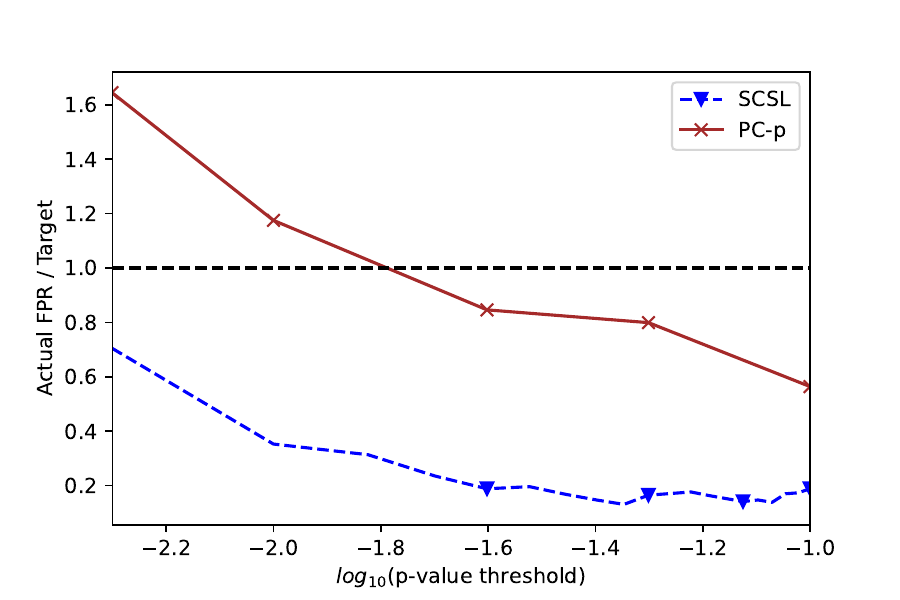}
      \includegraphics[width=0.44\linewidth]{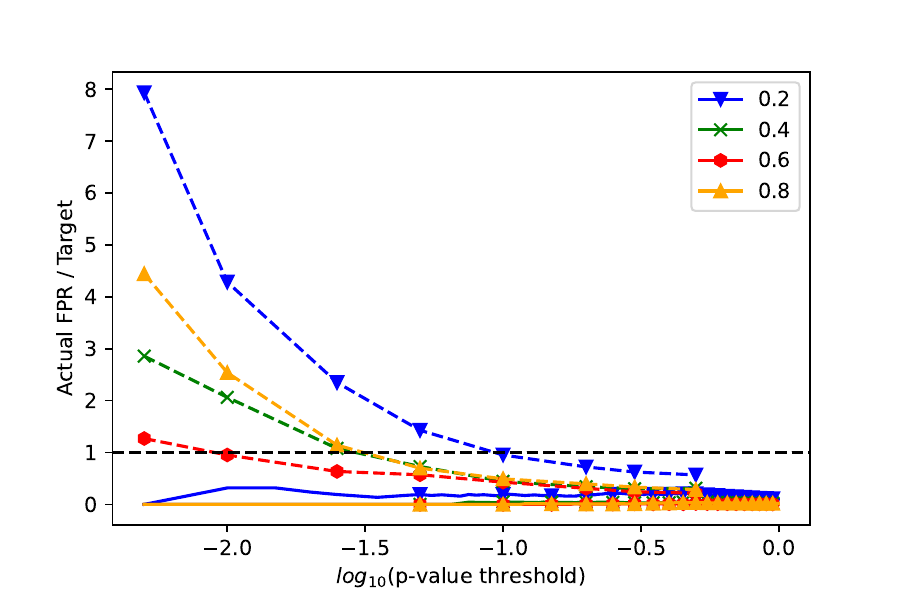}
\caption{Left hand side shows this ratio on a semi-synthetic dataset generated with real-world confounding which matches the complexity of the real dataset ($|\mathcal{X}|$ = 47 and $|\mathcal{Y}|$ = 23). Right hand side shows the ratio of actual false positive rate to target rate on the PC-p algorithm (dashed) compared with the proposed methodology (solid) on semi-synthetic datasets generated with synthetic confounding and confoundedness parameter $p \in \{0.2,0.4,0.6,0.8\}$.}
\label{fig:performance_collide}
\end{figure*}

\subsection{Semi-synthetic simulations with real-world confounding} \label{sec:semisynth_real}
In order to test the method, we need to produce a dataset that matches the structure of the actual data source, but with ground truth knowledge of the causal structure so that performance can be measured. To this end, we propose the construction of a semi-synthetic dataset. The method for data construction outlined in this section preserves the joint distribution of $\mathcal{X}$ and $\mathcal{Y}$, but allows for a conditional distribution $P(\mathcal{Y} \mid \mathcal{X})$ that is synthetic. 

\paragraph{Dataset construction} The algorithm takes the realized datasets $\mathcal{D}_{\mathcal{X}} \in \{0,1\}^{n \times p}$ and $\mathcal{D}_{\mathcal{Y}} \in \{0,1\}^{n \times m}$ and generates a synthetic data  for $\mathcal{Y}$ which we denote $\tilde{\mathcal{D}}_{\mathcal{Y}}$. To do this, we sample $K$ features from $\mathcal{X}$ for every element of $\mathcal{Y}$. We let $\mathcal{X}_{k}^\star$ be the features of $\mathcal{X}$ chosen for a particular $Y_{k}$ and sample coefficients $\beta_{k} \in \mathbb{R}^{K}$ from a $\mathcal{N}(2,1)$ distribution. 

Letting $\mathcal{D}_{\mathcal{X},k,i}^\star \in \{0,1\}^{ K}$ be the realized data points corresponding to the chosen features for the $i$th individual, we generate a logistic likelihood function $f_{Y_{k,i}}(\mathcal{D}_{\mathcal{X},k,i}^\star) := \frac{1}{1 +  \exp(-\beta_{k}^{T}\mathcal{D}_{\mathcal{X},k,i}^\star)}$.
An issue with using this likelihood function directly to generate the data is that the outcome dataset will no longer have the same dependence structure as the original. To circumvent this issue, we use the likelihood function to sample \emph{actual rows} of $\mathcal{D}_{\mathcal{Y}}$. Specifically, we generate $\tilde{\mathcal{D}}_{\mathcal{Y},i} \sim \text{Cat}( \theta_1,..., \theta_{n})$ where $\theta_i \propto \prod_k  f_{Y_{k,i}}(\mathcal{D}_{\mathcal{X},k,i}^\star)$. In other words, each row gets sampled in proportion to its overall likelihood under the assumed model. The procedure is described more explicitly in the Appendix. 

\begin{table} 
\centering
\begin{tabular}{| p{0.7cm}  p{0.4cm} p{0.4cm} |p{0.7cm} p{0.9cm}p{0.7cm} p{0.7cm} p{0.7cm} p{0.7cm} p{0.7cm} p{0.7cm} p{0.9cm} p{2cm}| }
\hline
\multirow{2}{*}[-2pt]{$n$}  & \multirow{2}{*}[-4pt]{$|\mathcal{X}|$}  &  \multirow{2}{*}[-4pt]{$|\mathcal{Y}|$}  &  \multicolumn{10}{|c|}{F1 Score}   \\
 & &  &  SCSL & PC-p & PC  & BOSS & CCD & FCI & FGES & 	GFCI & 	GRASP & 	GRaSP-FCI  
 \\
 \hline
200 & 5 & 5 & 	\textbf{0.26} & 	0.24 & 	 0.0 & 	 0.0 & 	 0.0 & 	 0.0 & 	 0.0 & 	 0.0 & 	 0.0 & 	 0.0 \\
&10 & 10 & 	0.07 & 	\textbf{0.10} & 	 0.0 & 	 0.0 & 	 0.0 & 	 0.0 & 	 0.0 & 	 0.0 & 	 0.0 & 	 0.0 \\
&15 & 15& 	\textbf{0.09} & 	0.07 & 	 0.0 & 	 0.0 & 	 0.0 & 	 0.0 & 	0.03 & 	0.03 & 	0.03 & 	0.06 \\
& 20& 20& 	0.04 & 	0.04 & 	0.02 & 	\textbf{0.11} & 	0.02 & 	0.02 & 	0.04 & 	0.04 & 	0.06 & 	0.06 \\
\hline
2000 & 5 & 5 & 	\textbf{0.71} & 	0.38 & 	 0.0 & 	0.18 & 	 0.0 & 	 0.0 & 	0.17 & 	0.17 & 	 0.0 & 	0.17 \\
&10 & 10 & 	\textbf{0.30} & 	0.14 & 	 0.0 & 	 0.0 & 	 0.0 & 	 0.0 & 	 0.0 & 	 0.0 & 	 0.0 & 	 0.0 \\
&15 & 15& 	\textbf{0.12} & 	0.10 & 	 0.0 & 	0.03 & 	 0.0 & 	 0.0 & 	 0.0 & 	  & 	 0.0 & 	0.03 \\
& 20& 20& 	\textbf{0.08} & 	0.06 & 	0.0 & 	0.04 & 	0.0 & 	0.0 & 	0.02 & 	 & 	0.04 & 	0.04 \\
 \hline
20,000 & 5 & 5 & 	0.87 & 	0.57 & 	\textbf{0.95} & 	0.84 & 	0.95 & 	0.82 & 	0.95 & 	0.89 & 	0.84 & 	0.89 \\
&10 & 10 & 	\textbf{0.78} & 	0.37 & 	0.29 & 	0.46 & 	0.29 & 	0.06 & 	0.46 & 	0.24 & 	0.38 & 	0.24 \\
&15 & 15& 	\textbf{0.49} & 	0.16 & 	 & 	0.15 & 	 & 	 & 	0.13 & 	0 & 	0.15 & 	0.06 \\
& 20& 20& 	\textbf{0.33} & 	0.06 & 	 & 	0.06 & 	 & 	 & 	0.04 & 	0.02 & 	0.08 & 	 \\

\hline
\end{tabular}
\caption{Comparison between F1 scores of SCSL comapred to existing methods for large-scale causal discovery for synthetic datasets constructed with real-world confounding. SCSL has an advantage over competing methods when the dimension of the node set ($\mathcal{X}, \mathcal{Y}$) is high relative to the sample size. Blank entries indicate the method failed to complete running after 12 hours of computation time.}
\label{tab:fes_fges}
\end{table}

\paragraph{Results} \label{sec:conditioning}
We first investigate how quickly the methods for optimizing the worst-case $p$-value described in \cref{sec:pval_opt} converge to the actual minimum, compared with the naive approach of randomly searching over the space in \cref{fig:performance_comp}. We see that the Gumbel-Softmax approach performs better than the naive approach at first, but then approaches an asymptote as it converges on a solution. The hybrid approach is able to dominate both approaches by allowing for both learning and exploration of the full combinatorial space. In this case, we manually choose to switch to sampling conditioning subsets without replacement at $200$ iterations, though in principle it may be possible to learn an optimal time to swap procedures from the data. 

\cref{fig:performance_collide} (left) shows the Type I error for SCSL and the PC-p algorithm on the generated data. We note that the SCSL algorithm achieves Type I error control, while the PC-p algorithm inflates the Type I error rate when $p$-values are small. To allow for comparison with methodologies that do not aim at frequentist error control, we also track the F1 score of SCSL along with benchmark methodologies in \cref{tab:fes_fges}. We note that SCSL outperforms existing methods when the dimension of the node set is large relative to the number of observations. Additional details about these experiments and more extensive empirical results are reported in the Appendix.

\subsection{Semi-synthetic simulations with synthetic confounding}   \label{sec:semisynth_synthetic}

Causal structure in $\mathcal{Y}$ is the key challenge in our task, as illustrated in \cref{fig:abxy-graph}. To evaluate the robustness of our method in the presence of different degrees of confoundedness, we construct a collection of semi-synthetic datasets where we stochastically control the degree of confounding, allowing us to stress test the methodology under adverse conditions. Compared with \cref{sec:semisynth_real}, these datasets have the disadvantage of not matching the structure of the actual dataset as closely. However, they are still a valuable point of comparison to assess performance across additional types of confoundedness. 

\paragraph{Dataset construction} We take as input the actual datasets $\mathcal{D}_{\mathcal{X}} \in \{0,1\}^{n \times p}$ and generates a synthetic dataset for $\mathcal{Y}$ which we label $\tilde{\mathcal{D}}_{y}$. For each desired $Y_{k}$, we choose $K$ features in $\mathcal{X}$ and generate coefficients $\beta_{k} \in \mathbb{R}^{K}$ sampled from a $\mathcal{N}(2,1)$ distribution. We store the chosen features in $\mathcal{X}_{Y_k}^\star$ and denote $\mathcal{D}_{\mathcal{X},k,i}^\star \in \{0,1\}^{K}$ to be the corresponding realized values of the chosen features for the $i$th individual. We now proceed by generating the elements of $\tilde{\mathcal{D}}_{y}$ which we call $Y_{k,i}$ in a sequential way. For the first element, the likelihood is defined simply as $f_{Y_{1,i}}(\mathcal{D}_{\mathcal{X},k,i}^\star) := \frac{1}{1 +  \exp(-\beta_{k}^{T}\mathcal{D}_{\mathcal{X},k,i}^\star)}$ and now $Y_{1,i} \sim \text{Ber}(f_{Y_{k,1}}(\mathcal{D}_{\mathcal{X},Y_1,i}^\star))$ is generated.  

For subsequent features $Y_{k,i}$, we also pick features $Y_{l}$ for $l < k$ with probability $p$ to contribute to the likelihood function. Denote $\mathcal{D}_{\mathcal{Y},Y_k,i}^\star$ as vector containing the chosen $Y_{l}$ for an individual $i$ and the coefficients, again sampled from a $\mathcal{N}(2,1)$ distribution, as $\gamma_{k}$. Now, define the likelihood as $f_{Y_{k,i}}(\mathcal{D}_{\mathcal{X},k,i}^\star, \mathcal{D}_{\mathcal{Y},k,i}^\star) := \frac{1}{1 +  \exp(-\beta_{k}^{T}\mathcal{D}_{\mathcal{X},k,i}^\star -\gamma_{k}^{T}\mathcal{D}_{\mathcal{Y},Y_k,i}^\star)  }$ and again sample $Y_{k,i} \sim \text{Ber}\left(f_{Y_{k,i}}(\mathcal{D}_{\mathcal{X},k,i}^\star, \mathcal{D}_{\mathcal{Y},Y_k,i}^\star)  \right)$. One can think of $p$ as a confounding parameter which makes the causal learning problem more difficult by increasing the number of dependencies in the graph. This algorithm is described in detail in the Appendix. 

\paragraph{Results} 
\cref{fig:performance_collide} (right) shows the Type I error control for our method and the PC-p algorithm across four different levels of confoundedness ($p \in \{0.2,0.4,0.6,0.8\}$). Across all four levels, the PC-p algorithm inflates the Type I error rate, sometimes by as much as 8x the nominal level. Comparisons in terms of F1 score and power are included in the Appendix, though we note results are broadly similar to those shown in \cref{tab:fes_fges} for the datasets constructed with real-world confounding. 

\subsection{Results on real dataset}

Finally, we run the methodology on the real data. We stratify the dataset based on the primary tumor site location, and calculate $p$-values separately within each stratum to identify the secondary tumor location and gene combinations that are significant across different metastases. A rejection set is formed using the Benjamini-Hochberg (BH) \citep{bh_procedure} procedure with target FDR of 0.05 and is shown in~\cref{tab:results_actual}. As a point of comparison, we also calculate the $p$-values for marginal tests of independence that the original paper used to identify connections. \cref{fig:pvalues_actual} compares the marginal $p$-values used by the original paper with the causal $p$-values calculated through SCSL. Applying the BH procedure with the same target threshold results in $161$ rejections of marginal $p$-values. However, only $6$ of these $161$ remained in the causal rejection set. 

The causal mutations discovered have mechanistic evidence in the biology literature. For instance, CDH1 has been mechanistically investigated in breast cancer models of metastasis through its connection to asparagine~\citep{knott2018asparagine}. Colon cancers are often treated with EGFR-inhibitors, which are ineffective in the presence of KRAS mutants which continue to activate the MAPK pathway, leading to eventual metastasis~\citep{prenen2010new}. TP53 mutations in certain pancreas cancers have been shown to increase fibrosis, enabling tumors to better evade the immune system and increasing metastatic potential~\citep{maddalena:etal:2021:pdac-tp53-mets}. While these mechanistic links between the primary site, gene, and general metastasis are known, the site-specific patterns have not been investigated. Thus, our causal testing results may provide valuable guidance to scientists and clinicians considering the utility of invasive patient monitoring.

  \begin{minipage}{\textwidth}
  \begin{minipage}[b]{0.4\textwidth}
    \centering
    \includegraphics[width= \linewidth]{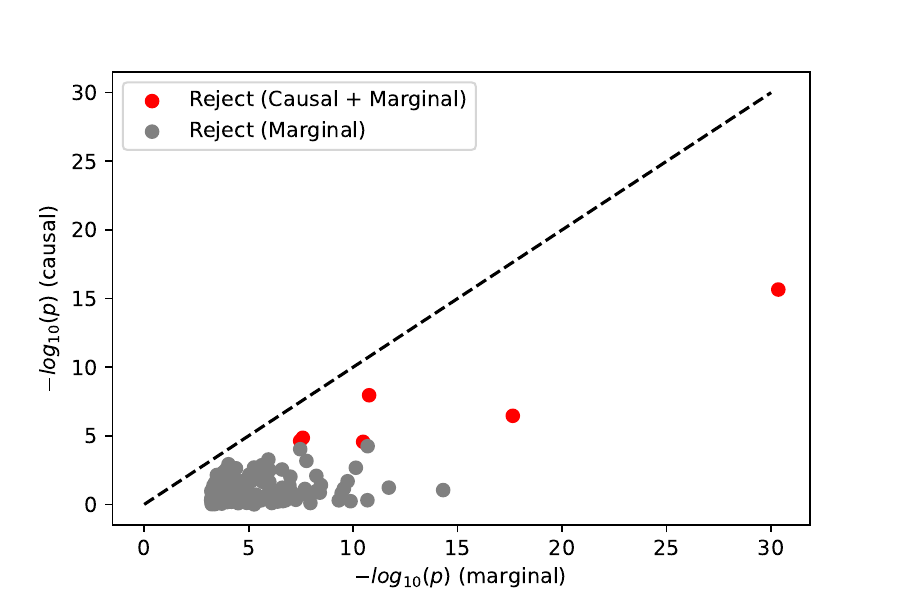}
    \captionof{figure}{Plot of the $161$ discoveries rejected using the BH procedure with target FDR of $0.05$ on the marginal independence associations. Only $6$ discoveries remain when applying the BH procedure with the same target threshold on the causal $p$-values.}
     \label{fig:pvalues_actual}
  \end{minipage}
  \hfill
  \begin{minipage}[b]{0.58\textwidth}
    \resizebox{1.0\textwidth}{!}{
\begin{tabular}{ |c c c | c c |}
\hline
&  &  & \multicolumn{2}{c|}{$p$-value}\\
Primary &  Gene & Secondary & Causal & Marginal \\
 \hline
Breast & CDH1 & Lung & $3.5 \times 10^{-7}$  & $2.3 \times 10^{-18} $ \\
Colon & KRAS & Lung & $1.4 \times 10^{-5}$  & $2.6 \times 10^{-8} $ \\
Liver & TERT & Liver & $2.3 \times 10^{-5}$  & $3.4 \times 10^{-8} $ \\
Lung & EGFR & CNS (Brain) & $2.8 \times 10^{-5}$  & $3.3 \times 10^{-11} $ \\
Pancreas & KRAS & Lymph & $2.2 \times 10^{-16}$  & $4.5 \times 10^{-31} $ \\
Pancreas & TP53 & Lymph & $1.1 \times 10^{-8}$  & $1.7 \times 10^{-11} $ \\


\hline
\end{tabular}}
      \captionof{table}{List of all $6$ genes and tumor combinations (corresponding to red markers on right graph) identified as significant at each primary site after forming a rejection set using the Benhamini-Hochberg procedure with target FDR of $0.05$. The causal $p$-values are significantly more conservative, leading to fewer rejections. }
      \label{tab:results_actual}
    \end{minipage}
  \end{minipage}

%% file: CLEAR_2025/conclusion.tex
\section{Conclusion}
We introduced a new algorithm for causal discovery which drastically decreases the computational burden required to compute a $p$-value for a causal relationship between two nodes in a directed acyclic graph with temporally separated sets of variables. We tested this methodology on semi-synthetic data constructed from a recent study on somatic tumor mutations and metastatic potential for a panel of patients and found that the methodology successfully controlled Type I error and had reasonable power across datasets of differing levels of confoundedness. When run on the dataset of \cite{data_source} 
, interesting connections between metastases and genes are identified. 

Several avenues for follow-up work exist. From a statistical perspective, the $p$-values generated from the procedure are conservative and power can potentially be improved through post-hoc adjustment to the $p$-values, for example through the use of Empirical Bayes methods \citep{efron}. From a computational perspective, more sophisticated probability models could be used to find the worst-case $p$-value to search the combinatorial space. 

Other areas of improvement relate to the causal ordering assumption that edges can only be directed from $\mathcal{X}$ to $\mathcal{Y}$ due to separation in time. For datasets that cannot be partitioned in this way, many aspects of the method can still be used to improve computational and statistical efficiency such as $p$-value optimization and amortized predictive modeling. In addition to increasing the computational burden of the method, this would leave open the question of how to orient the edges after skeleton discovery when used on more generic datasets. Alternatively, instead of only two groups of temporally seperated nodes, it would be interesting to investigate how this methodology performs when adapted to datasets with several groups of nodes coming into existence over time. Finally, although we have motivated our method from a metastasis dataset, the methodology is general and could be applied to a number of datasets in areas like genome-wide association studies.

%% file: CLEAR_2025/appendix.tex
\section{Omitted Proofs}

\paragraph{Proof of Proposition~\ref{prop:edge}} 

First, assume that there is an edge between $X_{j}$ and $Y_{k}$. Then, $X_{j}$ and $Y_{k}$ are not d-separated given any set of nodes $V \subseteq X_{-j} \cup Y_{-k}$. By Assumption~\ref{assumption:dsep}, $X_{j}$ and $Y_{k}$ will be conditionally dependent given $V$. In particular, they will be conditionally dependent for $S \cup X_{-j}$ for all $S \subseteq Y_{-k}$.

Next, assume that $X_{j}$ and $Y_{k}$ are conditionally dependent given $S \cup X_{-j}$ for all $S \subseteq Y_{-k}$. Since no edges are directed from any element in $\mathcal{Y}$ to any element in $\mathcal{X}$, then there are no colliders in $W$ for any $W \subseteq X_{-j}$. This implies that $X_{j}$ and $Y_{k}$ are also conditionally dependent given $T \subseteq X_{-j} \cup Y_{-k}$. This implies that $X_{j}$ and $Y_{k}$ are not $d$-separated given any $T$. \Cref{assumption:markov} and \cref{assumption:sufficiency} then together imply that $X_{j}$ and $Y_{k}$ must share an edge. 

\section{Algorithms for construction of semi-synthetic datasets}
We include the pseudocode for construction of semi-synthetic datasets in \cref{alg:collide_synth} and \cref{alg:semi_synth} below. 

\begin{algorithm}[htb] 
   \caption{Semi-synthetic dataset with real-world confounding}\label{alg:semi_synth}
   \begin{algorithmic}
   \STATE {\bfseries Input:} Data: $\mathcal{D}_{\mathcal{X}} \in \{0,1\}^{n \times p}$ and $\mathcal{D}_{\mathcal{Y}} \in \{0,1\}^{n \times m}$, 
   \STATE Shuffle rows of $\mathcal{D}_{\mathcal{X}}$ and store as $\tilde{\mathcal{D}}_{\mathcal{X}}$
   \STATE Shuffle rows of $\mathcal{D}_{\mathcal{Y}}$ 
   \STATE Sample $K$ features from $\mathcal{X}$ for each element of $\mathcal{Y}$  and denote $\mathcal{X}_{k}^\star$ the features for a particular $Y_{k}$
   \STATE Create likelihoods for the rows of $\mathcal{Y}$ from  $\mathcal{X}^\star$ by 
    \begin{enumerate}\setlength\itemsep{-3pt}
        \item Sampling coefficients $\beta_{ k} \in \mathbb{R}^{K}$ for $\mathcal{X}_{Y_k}^\star$  (e.g., from a standard normal)
        \item Let $f_{Y_{k,i}}(\mathcal{D}_{\mathcal{X},k,i}^\star) := \frac{1}{1 +  \exp(-\beta_{k}^{T}\mathcal{D}_{\mathcal{X},k,i}^\star)}$ be the likelihood associated with each response $Y_k$ and row $i$, where $\mathcal{D}_{\mathcal{X},k,i}^\star \in \{0,1\}^{ K}$ are the realized data corresponding to $\mathcal{X}_{k}^\star$ for the $i$th individual in the dataset.
    \end{enumerate}
    \FOR{row $i=1$ {\bfseries to} $n$}
        \STATE sample ${\tilde{\mathcal{D}}}_{\mathcal{Y},i}\sim \text{Cat}(\theta_1, \dots, \theta_n)$ with $\theta_i \propto \prod_k f_{Y_{k,i}}(\mathcal{D}_{\mathcal{X},k,i}^\star)$
    \ENDFOR
    \STATE {\bfseries Return} semi-synthetic data $\tilde{\mathcal{D}}_{\mathcal{X}}$ and $\tilde{\mathcal{D}}_{\mathcal{Y}}$ 
    \end{algorithmic}
\end{algorithm}

\begin{algorithm}[htb] 
   \caption{Semi-synthetic dataset with synthetic confounding} \label{alg:collide_synth}
   \begin{algorithmic}
    \STATE Data: $\mathcal{D}_{\mathcal{X}} \in \{0,1\}^{n \times p}$
   \STATE Shuffle rows of $\mathcal{D}_{\mathcal{X}}$ and store as $\tilde{\mathcal{D}}_{\mathcal{X}}$
   \FOR{$k=1$ {\bfseries to} $|\mathcal{Y}|$}
       \item Sample $K$ features from $\mathcal{X}$ and denote $\mathcal{X}_{k}^\star$ the features for a particular $Y_{k}$ 
       \item For each elements $Y_{k} \in \mathcal{S}_{2}$ with $l < k$, add $Y_{l}$ to $\mathcal{Y}_{Y_{k}}^\star$ with probability $p$.
       \item Sample coefficients $\beta_{k} \in \mathbb{R}^{K}$ for $\mathcal{X}_{Y_k}^\star$ (e.g., from a standard normal) and $\gamma_{k} \in \mathbb{R}^K$ for $\mathcal{Y}_{k}^\star$.
       \item Let $f_{Y_{k,i}}(\mathcal{D}_{\mathcal{X},k,i}^\star, \mathcal{D}_{\mathcal{Y},k,i}^\star) := \frac{1}{1 +  \exp(-\beta_{k}^{T}\mathcal{D}_{\mathcal{X},k,i}^\star -\gamma_{k}^{T}\mathcal{D}_{\mathcal{Y},k,i}^\star)  }$ the likelihood associated with each response $Y_k$, where $\mathcal{D}_{\mathcal{X},k,i}^\star$ is the realized data corresponding to $\mathcal{X}_{k}^\star$ and $\mathcal{D}_{\mathcal{Y},k,i}^\star$ is a vector corresponding to the realized data for $\mathcal{Y}_{k}^\star$ for the $i$th individual in the dataset
    \FOR{row $i=1$ {\bfseries to} $n$}
        \STATE sample ${\tilde{\mathcal{D}}}_{Y_{k},i} \sim \text{Ber}(\theta_{i,k}) $ with $\theta_{i,k} = f_{Y_{k,i}}(\mathcal{D}_{\mathcal{X},Y_k,i}^\star, \mathcal{D}_{\mathcal{Y},Y_k,i}^\star)$
    \ENDFOR
    \ENDFOR
    \STATE {\bfseries Return} semi-synthetic data $\tilde{\mathcal{D}}_{\mathcal{X}}$ and $\tilde{\mathcal{D}}_{\mathcal{Y}}$ 
    \end{algorithmic}
\end{algorithm}

\section{Adjustments for continuous and mixed-value data}  \label{subsec:continuous_meth}
In \cref{alg:batch_grad_continuous}, we present a slightly augmented the methodology to train amortized predictive models to accommodate continuous valued data. The algorithm is largely the same as the one presented in the main paper, with the modification that \emph{both} the masks $(B,M)$ and the interactions between the mask are inputs into the loss function. In the case that binary-valued data was used with $X_{j},Y_{k} \in \{-1,1\}$, then this was not necessary as the interactions would simply lead to the masked versions $\tilde{X}_{j},\tilde{Y}_{k} \in \{-1,0,1\}$. In the case of continuous or mixed valued data, however, there is a need to distinguish between data that is $0$ because it has been masked or data that is \emph{actually} $0$. 

To construct semi-synthetic datasets to test this dataset, \cref{alg:semi_synth} and \cref{alg:collide_synth} can be modified slightly by letting the likelihood functions correspond to a Gaussian distribution instead of a logistic response, and then sampling appropriately.   
\begin{algorithm}[tb]
   \caption{Amortized predictive model training for continuous response data} \label{alg:batch_grad_continuous}
\begin{algorithmic}
   \STATE {\bfseries Input:} Data: $\mathcal{D}_{\mathcal{X}} \in \mathbb{R}^{n \times |\mathcal{X}|}$; $\mathcal{D}_{\mathcal{Y}} \in \mathbb{R}^{n \times |\mathcal{Y}|}$; $n_{ep}$ (number of epochs), $n_{batch}$ (batch size), $p$ (masking parameter), loss function $\ell \left(\theta, \mathcal{X}, Y_{-k}, Y_{k},B,M \right)$
    \FOR{$i=1$ {\bfseries to} $n_{ep}$}
      \FOR{$i=1$ {\bfseries to} $[n/n_{batch}]$}
       \STATE Draw $n_{batch}$ new samples
       \STATE Draw $B_{m} \sim \text{Ber}(p)$ for each $Y_{m} \in Y_{-k}$. Denote $B = \left(B_{1},\cdots, B_{k-1},B_{k+1},...,B_{|\mathcal{Y}|}\right)$.
       \STATE Draw $M \sim\text{Cat}(\mathcal{X},q)$ with equally-weighted probabilities
       \STATE Construct $\tilde{Y}_{-k}$ by taking each $Y^{i}_{m}$ in the sample and replacing with $\widetilde{Y}^{i}_{m} := Y^{i}_{m} \times B_{m}$
       \STATE Construct $\tilde{\mathcal{X}}$ by taking each $X^{i}_{j}$ and replacing with $\widetilde{X}^{i}_{j} := X^{i}_{j} \times \mathbbm{1}_{M \ne k}$
       \STATE Compute $\frac{\partial \ell(\tilde{\theta,\mathcal{X}}, \tilde{Y}_{-k}, Y_{k},B,M)}{\partial \theta}$
       \STATE $\theta \leftarrow \theta - \eta  \frac{\partial \ell(\theta,\tilde{\mathcal{X}}, \tilde{Y}_{-k}, Y_{k},B,M)}{\partial \theta}$ 
    \ENDFOR
   \ENDFOR
\end{algorithmic}
\end{algorithm}


\section{Flowchart summarizing methodology}
In \cref{fig:flowchart}, we present a visual flowchart illustrating how to apply the methodology. Assumed inputs into this process are a method for computing $p$-values corresponding to conditional independence tests. For all of our experiments, we use amortization to efficiently compute the GCM test statistic  for different conditioning subsets. 

\tikzstyle{default} = [rectangle, rounded corners,text centered,draw=black,minimum height = 0.75cm,minimum width = 3.5cm,align=center]

\tikzstyle{annotation} = [text centered,align=center]

\tikzstyle{arrow} = [thick,->,>=stealth]

\begin{figure}[ht] 
    \centering
\begin{tikzpicture}
\node (step_1) [default,minimum width = 0.5cm] {Sample \\ $S^{t} \sim \text{Ber}(\theta_{1}^{t}) \times  \text{Ber}(\theta_{p}^{t}) $};
\node (step_1a) [default,right = 0.3cm of step_1,minimum width = 1cm] {Add $S^{t}$ to $S_{\text{list}}$};
\node (step_2) [default, below = 0.3cm of step_1] {Calculate $T^{(n)}(S^{t})$};
\node (step_3) [default, below = 0.3cm of step_2,fill=blue!20] {Compute $p$-value $\hat{p}^{t}$};
\node (step_3a) [default, right = 0.3cm of step_3,fill=blue!20,minimum width = 1cm] {Add $\hat{p}^{t}$ to $p_{\text{list}}$};
\node (step_3b) [annotation, below left = 0.3cm and 0cm  of step_3] {$\hat{p}^{t} \le \alpha$};
\node (step_3c) [annotation, below right = 0.3cm and 0cm  of step_3] {$\hat{p}^{t} > \alpha$};
\node (step_4) [default,fill=yellow!20, below  = 0.5cm  of step_3c] {End search; \\ set $\hat{p}_{X_{j} \rightarrow Y_{k}} := 1$ };
\node (step_5) [default, below  = 0.5cm  of step_3b] {Update \\ $\theta^{t+1} = \theta^{t} - \gamma \frac{\partial T^{(n)}}{\partial \theta^{t}}$  };
\node (step_5a) [annotation, below left = 0.3cm and 0cm  of step_5] {$t+1 \le q_1$};
\node (step_5b) [annotation, below right = 0.3cm and 0cm  of step_5] {$t+ 1 > q_1$};
\node (step_6) [default, right = 0.3cm of step_5b] {Compute $w_{S} = \prod_{i=1}^{p} [\theta_{i}^{t}]^{\mathbbm{1}_{i \in S}} [1 - \theta_{i}^{t}]^{\mathbbm{1}_{i \not\in S}} $ \\ for every $S \in \mathcal{P} (\mathcal{S}_{2} \ Y_{k}) \setminus S_{\text{list}}$};
\node (step_7) [default, below = 0.5 cm of step_6] {Draw $S^{t} \sim \text{Cat}(\pi_{1},...,\pi_{L})$ with \\ $\pi_{S} \propto w_{S}$ for every $S \in \mathcal{P} (\mathcal{S}_{2} \ Y_{k}) \setminus S_{\text{list}}$};
\node (step_7a) [default,right = 0.3cm of step_7,minimum width = 1cm] {Add $S^{t}$ to $S_{\text{list}}$};
\node (step_8) [default, below = 0.3cm of step_7] {Calculate $T^{(n)}(S^{t})$};
\node (step_9) [default, below = 0.3cm of step_8,fill=blue!20] {Compute $p$-value $\hat{p}^{t}$};
\node (step_9a) [default, right = 0.3cm of step_9,fill=blue!20,minimum width = 1cm] {Add $\hat{p}^{t}$ to $p_{\text{list}}$};
\node (step_9b) [annotation, below left = 0.3cm and 0cm  of step_9] {$\hat{p}^{t} \le \alpha$};
\node (step_9c) [annotation, below right = 0.3cm and 0cm  of step_9] {$\hat{p}^{t} > \alpha$};
\node (step_10) [default,fill=yellow!20, below right = 0.5cm and 0cm of step_9c] {End search; \\ set $\hat{p}_{X_{j} \rightarrow Y_{k}} := 1$ };
\node (step_10a) [annotation, below left = 0.3cm and 0cm  of step_9b] {$t+1 \le q_1 +q_{2}$};
\node (step_10b) [annotation, below right = 0.3cm and 0cm  of step_9b] {$t+1 \ge q_1 + q_2$};
\node (step_11) [default,fill=yellow!20, below  = 0.5cm and 0cm of step_10b] {End search; \\ set $\hat{p}_{X_{j} \rightarrow Y_{k}} := \text{max} p_{\text{list}}$ };

\node(legend_1) [default,fill=yellow!20,minimum height = 0.5cm,minimum width = 0.75cm,
below left  = 1.5cm  and 0.7cm of step_5,label=east:``Inner'' p-value  ] {};
\node(legend_2) [default,fill=blue!20,minimum height = 0.5cm,minimum width = 0.75cm,
below left  = 2.25cm  and 0.7cm of step_5,label=east:``Outer'' p-value ] {};
\node(legend_3) [default,fill=white!20,draw=white,minimum height = 0.5cm,minimum width = 0.75cm,
below left  = 3cm  and 0.7cm of step_5,label=east:Cutoff for early stopping rule] {$\alpha$};
\node(legend_4) [default,fill=white!20,draw=white,minimum height = 0.5cm,minimum width = 0.75cm,
below left  = 3.75cm  and 0.7cm of step_5,label=east:No. iterations to train $\theta$] {$q_{1}$};
\node(legend_5) [default,fill=white!20,draw=white,minimum height = 0.5cm,minimum width = 0.75cm,
below left  = 4.5cm  and 0.7cm of step_5,label=east:No. iterations for search process] {$q_{2}$};

\draw [arrow] (step_1) -- (step_2);

\draw [arrow] (step_1) -- (step_1a);
\draw [arrow] (step_2) -- (step_3);
\draw [arrow] (step_3) -- (step_3a);

\draw [arrow] (step_3) -- (step_3b);

\draw [arrow] (step_3) -- (step_3c);
\draw [arrow] (step_3c) -- (step_4);

\draw [arrow] (step_3b) -- (step_5);
\draw [arrow] (step_5) -- (step_5a);
\draw [arrow] (step_5) -- (step_5b);
\draw [arrow] (step_5a) to [out=90,in=180]  (step_1);
\draw [arrow] (step_5b) -- (step_6);
\draw [arrow] (step_6) -- (step_7);
\draw [arrow] (step_7) -- (step_7a);
\draw [arrow] (step_7) -- (step_8);
\draw [arrow] (step_8) -- (step_9);
\draw [arrow] (step_9) -- (step_9a);
\draw [arrow] (step_9) -- (step_9b);
\draw [arrow] (step_9) -- (step_9c);
\draw [arrow] (step_9c) -- (step_10);

\draw [arrow] (step_9b) -- (step_10a);

\draw [arrow] (step_9b) -- (step_10b);
\draw [arrow] (step_10a) to [out=90,in=180]  (step_7);

\draw [arrow] (step_10b) -- (step_11);
\end{tikzpicture}
\caption{Flowchart illustrating the methodology. Nodes colored in blue represent $p$-values for the individual conditional independence tests corresponding to the ``inner'' null that $X_{j} \ind Y_{k} | S, X_{-j}$. Nodes colored in yellow correspond to $p$-values testing the ``outer'' null that $X_{j} \rightarrow Y_{k}$ is absent.}\label{fig:flowchart}

\end{figure}
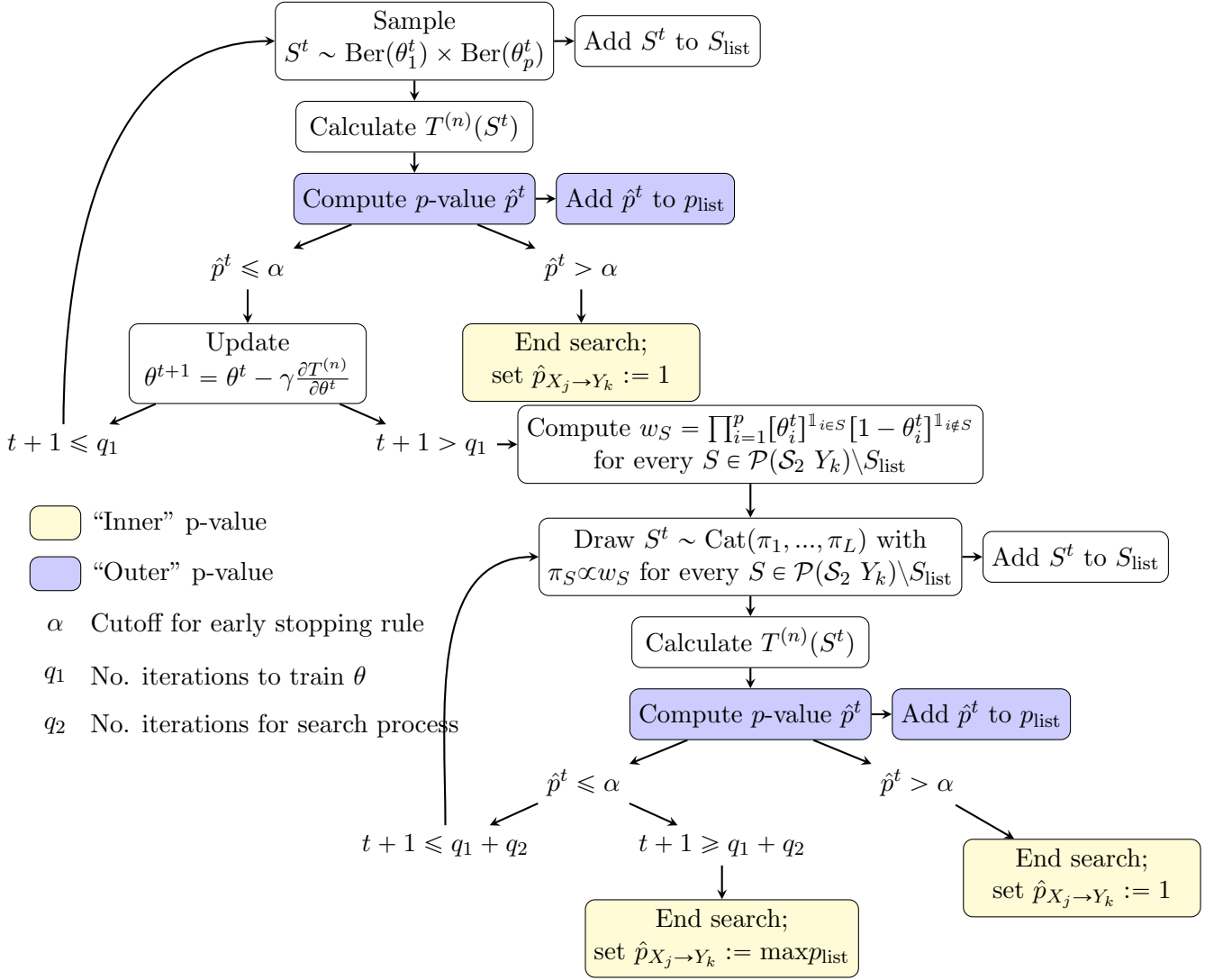

\section{Technical exposition of Generalized Covariance Measure} \label{sec:appendix_GCM}
We recall the details of the technical conditions described in \cite{hardness_CI} for completeness. We assume that the dataset consists of i.i.d. $n$ samples, with individual observations denoted $\{X_{j}^{i}, Y_{k}^{i}, S^{i}, X_{-j}^{i}\}_{i=1}^{n}$. Let $\mathcal{P}_{0}$ denote the family of joint distributions corresponding to the null that $X_{j} \ind Y_{k} | S, X_{-j}$.

Define $R_{i} = \left( X_{j}^{i} - \widehat{X}_{j}^{i} \right) \left( Y_{k}^{i} - \widehat{Y}_{k}^{i} \right)$ and the test statistic:
\begin{equation} 
    T^{(n)}=\frac{\sqrt{n} \cdot \frac{1}{n} \sum_{i=1}^n R_i}{\left(\frac{1}{n} \sum_{i=1}^n R_i^2-\left(\frac{1}{n} \sum_{r=1}^n R_r\right)^2\right)^{1 / 2}}.
\end{equation}
Further denote $\epsilon_{j,P} := X_{j} - \mathbb{E}_{P}[X_{j} | S, X_{-j}] $ and $\epsilon_{k,P} := Y_{k} - \mathbb{E}_{P}[Y_{k} | S, X_{-j}] $.

\begin{fact}[Theorem 6 of \cite{hardness_CI}] Define the following quantities:
$$A_f :=\frac{1}{n} \sum_{i=1}^n\left\{X_{j}^{i} - \widehat{X}_{j}^{i} \right\}^2, \text{  } A_g :=\frac{1}{n} \sum_{i=1}^n\left\{Y_{k}^{i} - \widehat{Y}_{k}^{i}\right\}^2,$$
$$B_f :=\frac{1}{n} \sum_{i=1}^n\left\{X_{j}^{i} - \widehat{X}_{j}^{i}\right\}^2
\mathbb{E}_{P} \left( \epsilon_{j,P}^{2} | S, X_{-j} \right), \text{ and } B_g :=\frac{1}{n} \sum_{i=1}^n\left\{Y_{k}^{i} - \widehat{Y}_{k}^{i}\right\}^2 \mathbb{E}_{P} \left( \epsilon_{k,P}^{2} | S, X_{-j} \right) .$$
\begin{enumerate}
    \item If for $P \in \mathcal{P}_{0}$, $A_{f} A_{g} = o(n^{-1})$, $B_{f} = o(1)$, $B_{g} = o(1)$, and $0 \le \mathbb{E}_{P} \left( \epsilon_{j,P}^{2} \epsilon_{k,P}^{2} \right)$, then  $\sup_{t \in \mathbb{R}} | \mathbb{P}_{P} \left(  T^{(n)} \le t \right) -  \Phi(t) | \rightarrow 0$.
    \item Let $\mathcal{P} \subseteq \mathcal{P}_{0}$ denote the set of  null distributions such that  $A_{f} A_{g} = o(n^{-1})$, $B_{f} = o(1)$, $B_{g} = o(1)$, $\inf_{\mathcal{P} \in P} \mathbb{E}_{p} \mathbb{E}_{P} \left( \epsilon_{j}^{2} \epsilon_{k}^{2} \right) \ge c_{1}$, and $\sup_{\mathcal{P} \in P} \mathbb{E}_{p} \mathbb{E}_{P} \left( \epsilon_{j,P}^{2} \epsilon_{k,P}^{2} \right) \le c_{2}$ for some $c_{1}, c_{2} >0$. Then $\sup_{P \in \mathcal{P}}\sup_{t \in \mathbb{R}} | \mathbb{P}_{P} \left(  T^{(n)} \le t \right) -  \Phi(t) | \rightarrow 0$.
\end{enumerate}
\end{fact}

\section{Additional experimental results} \label{appendix:semi}
Additional results for each of the semi-synthetic simulations are shown in this section. 

\begin{figure}[H]
    \centering
        \includegraphics[width=0.48\linewidth]{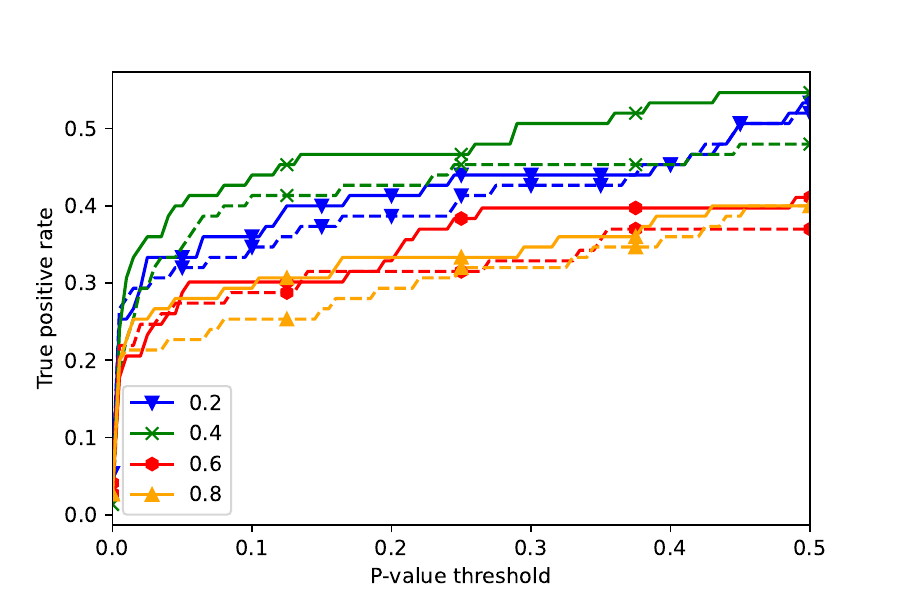}
        \includegraphics[width=0.48\linewidth]{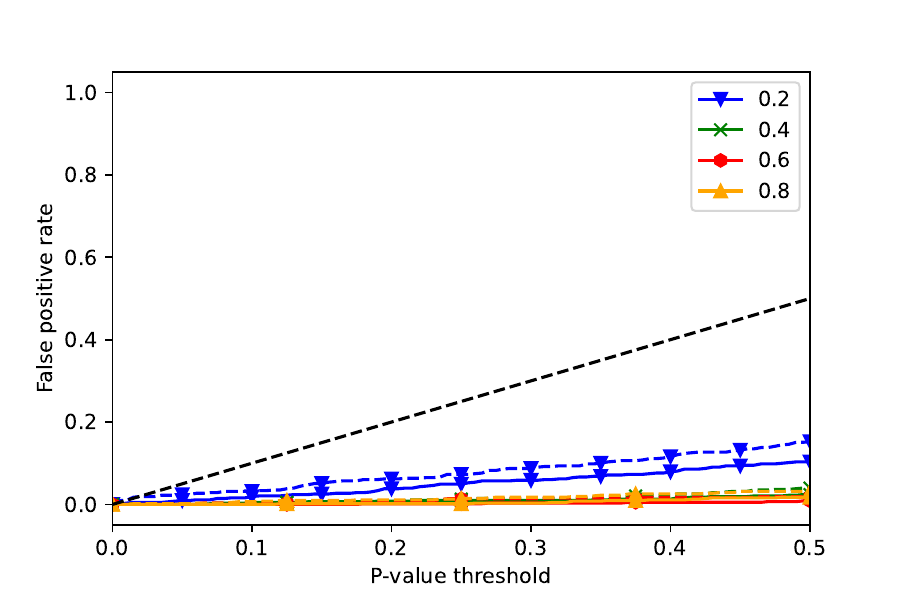}
         \includegraphics[width=0.48\linewidth]{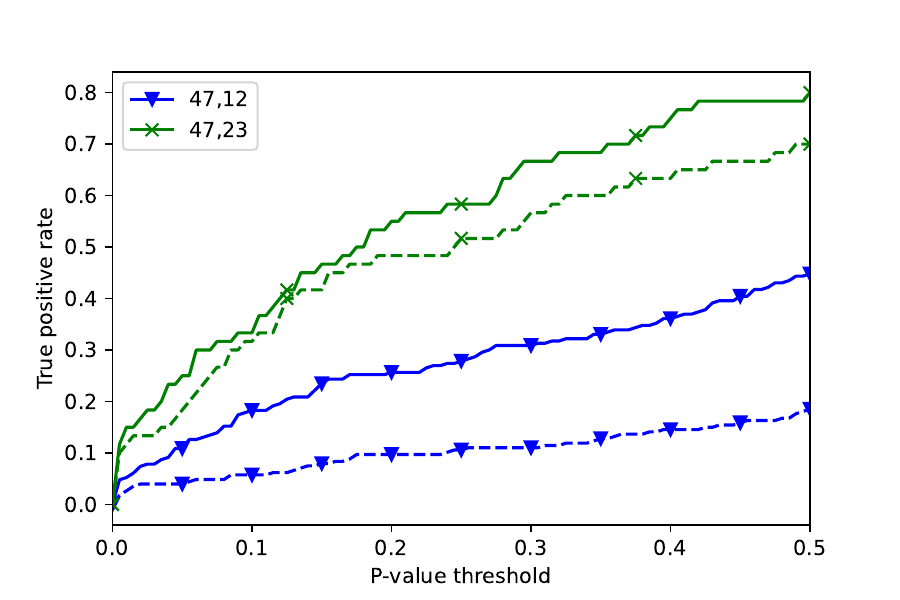}
        \includegraphics[width=0.48\linewidth]{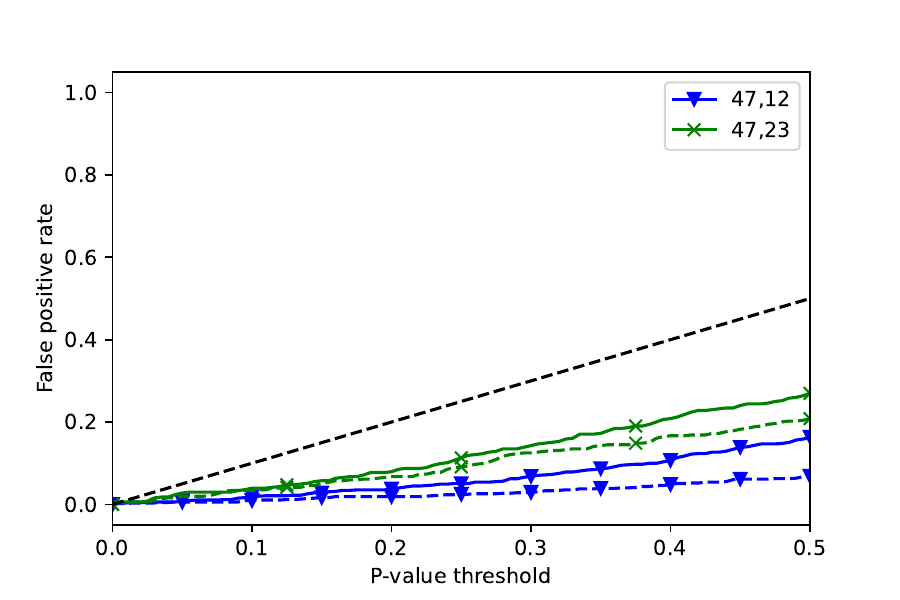}
    \caption{Type I error and Type II error rates for SCSL compared when logistic models (solid) are swapped out for a neural network (dashed). The top row shows results for the datasets with synthetic confounding while the bottom row shows results for datasets constructed with real confounding (note that the tuple $(a,b)$ in the legend indicates the number of variables contained in each node set with $|\mathcal{X}| = a$ and $|\mathcal{Y}|=b$ respectively). The performance across all methods is comparable, though we note the logistic model has higher power in general.}\label{fig:model_comparisons}
\end{figure}

\paragraph{Additional information on alternative predictive modelling approaches} In addition to using a logistic regression to compute the underlying regression functions in the GCM test statistic, we also experimented with a multi-layer perceptron neural network with with $2$ hidden layers comprised of $200$ nodes in each layer and dropout regularization. The performance of this method is compared and contrasted with the logistic regression used in the main paper in \cref{fig:model_comparisons}. Since the logistic regression had higher power, we focused on using it as a predictive model in the main paper.

\begin{figure}[H]
\centering
    \includegraphics[width=0.48\linewidth]{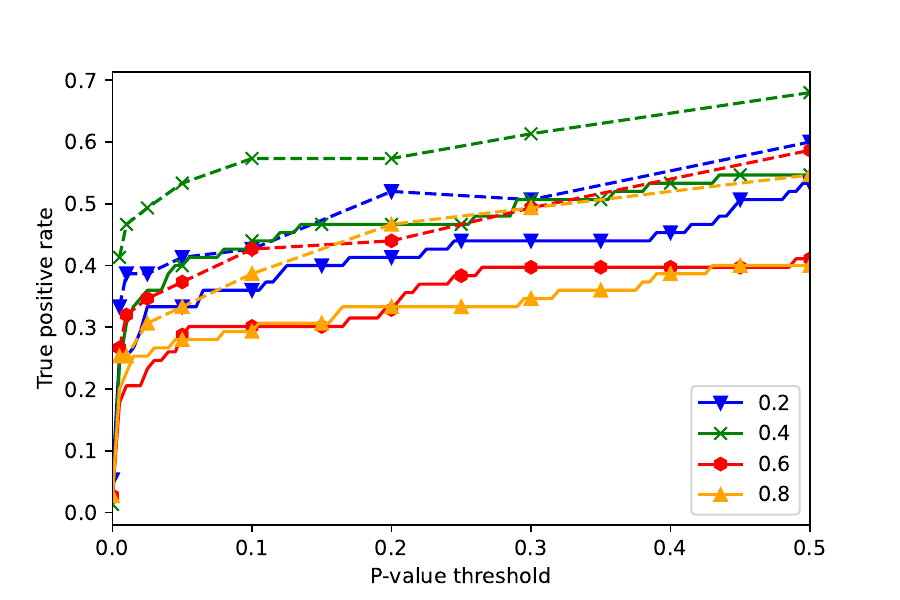}
    \includegraphics[width=0.48\linewidth]{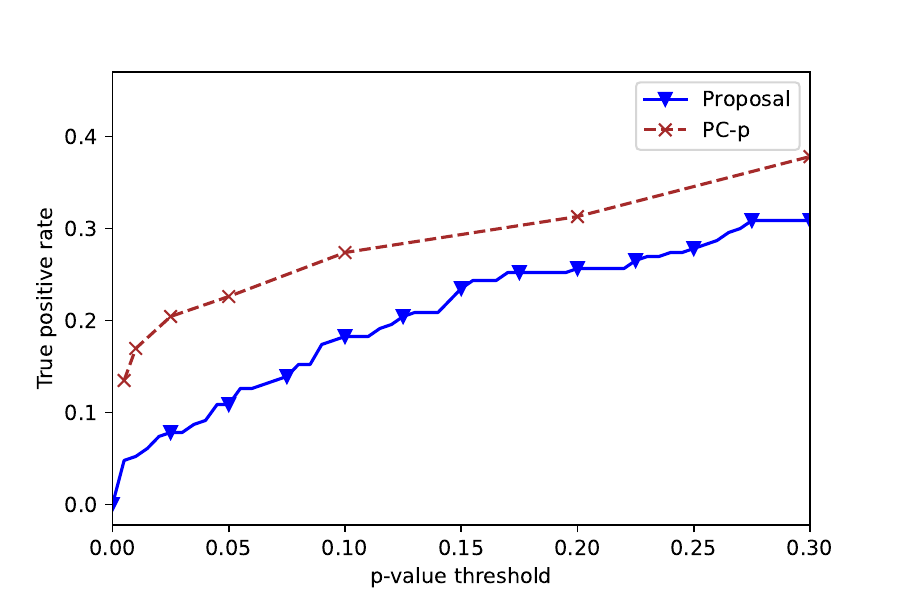}
    \caption{True positive rate for proposed methodology (solid) contrasted with PC-p method (dashed). Left-hand side shows the power for each of these methodologies run on datasets constructed by \cref{alg:collide_synth} for the same sequence of confoundedness parameters discussed in the main paper with $p \in \{0.2,0.4,0.6,0.8 \}$. Right-hand side shows the power for these two methods on the semi-synthetic dataset constructed with \cref{alg:semi_synth}.}
    \label{fig:benchmarks_power}
\end{figure}

\paragraph{Additional results on power} As discussed in the results section, the PC-p algorithm does have higher power than SCSL, but this comes at the expense of Type I error control. This is a natural consequence of the fact that the PC-p algorithm draws an edge between two nodes based on a consideration set of CI tests that is a subset of the tests considered by SCSL. Since we use the same underlying CI test in our comparison of the methodologies, this necessitates that the PC-p algorithm will have strictly lower $p$-values than SCSL. Nonetheless, the decrease in power resulting from a move to this new methodology is not significant. We compare the power of the two methodologies side by side in \cref{fig:benchmarks_power} for the same configuration of datasets discussed in the results section.

\paragraph{Comparisons to procedures without frequentist error guarantees}

Although the PC-p algorithm is the only existing method that aims at constructing $p$-values with frequentist error control, we also benchmark this procedure against approaches not aimed at controlling type I error that are implemented in the \textsc{TETRAD} project. The only two methods that scale reasonably to datasets of a similar size as the dataset of \cite{data_source} are Fast Greedy Equivalence Search (FGES) \citep{fges} and the Best Order Score Search (BOSS) algorithm \citep{greedy_sparest}. Other methods used as baselines are Cyclic Causal Discovery (CCD, \cite{CCD}) , Fast Causal Inference (FCI, \cite{FCI}), Greedy Fast Causal Inference (GFCI, \cite{gfci}), Greedy relaxation of the sparsest permutation (GRaSP, \cite{greedy_sparest}), and GRaSP-FCI \citep{gfci}. 

We note that above methods produce output in the form of a learned graph rather than a numeric $p$-value. To facilitate comparisons, we compare the accuracy of the causal graph learned via these methods with a causal graph produced by taking the $p$-values from the SCSL and PC-p methods and drawing an edge when the $p$-value is below $0.1$. We summarize the accuracy of these methods (in terms of ability to detect edges when present) using F1 score. The F1 scores are shown in \cref{tab:discrete_f1} and overall wall time is shown in \cref{tab:discrete_duration}. The experiments are conducted across variety of dataset configurations described in this paper: synthetic versus real-world confounding, degree of confoundedness ($p$), size of dataset, and size of node set. A similar set of results for continuous-valued datasets constructed using the methods in \cref{subsec:continuous_meth} are shown in \cref{tab:continuous_f1} and \cref{tab:continuous_duration}.

We note that although the primary strength of SCSL is its ability to control the error rate by producing valid $p$-values, the experimental results demonstrate that the method is also competitive with existing methods in terms of accuracy, power, and speed of computation. SCSL often has the highest F1 score among the tested methods and is nearly as fast as the BOSS and FGES algorithms.

\begin{figure}[H] 
    \centering
        \includegraphics[width=0.32\linewidth]{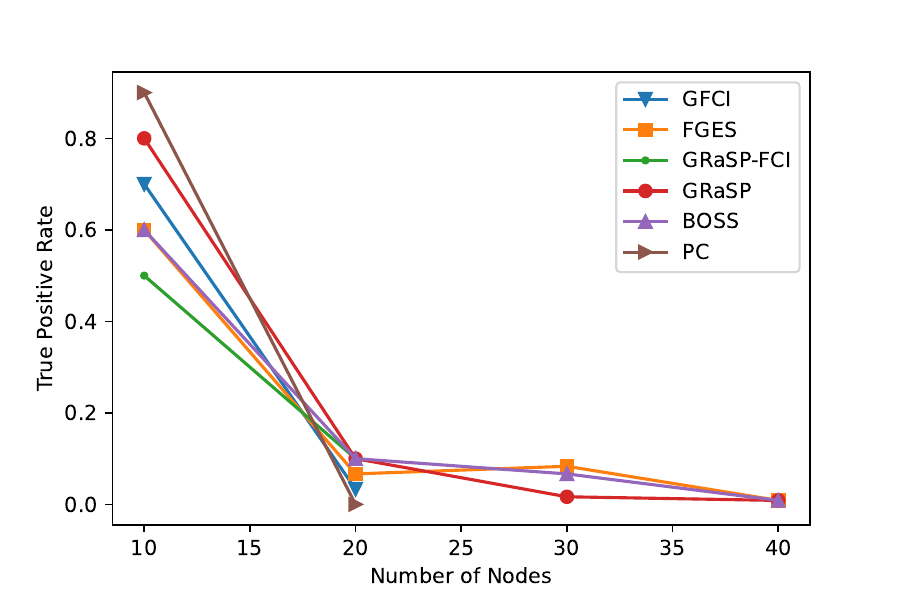}
        \includegraphics[width=0.32\linewidth]{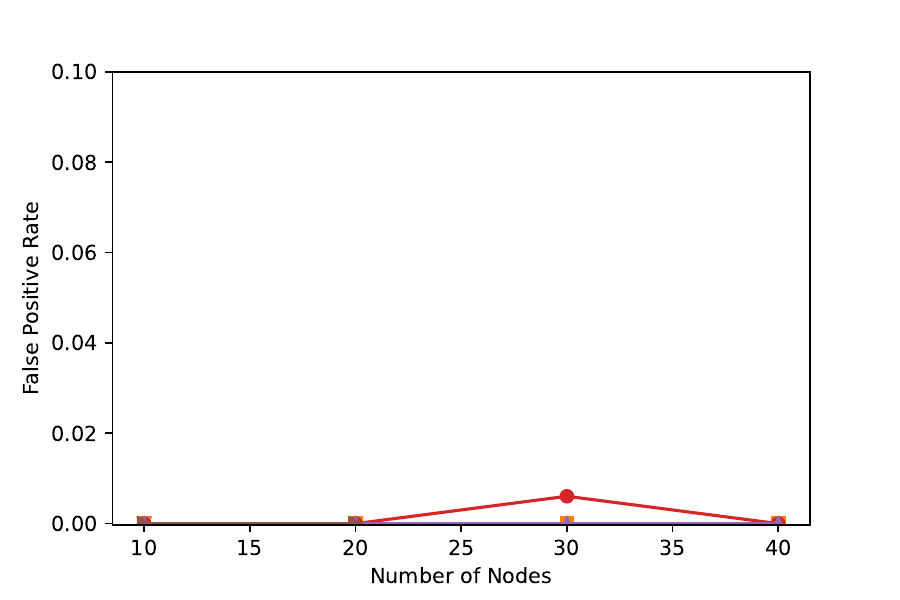}
         \includegraphics[width=0.32\linewidth]{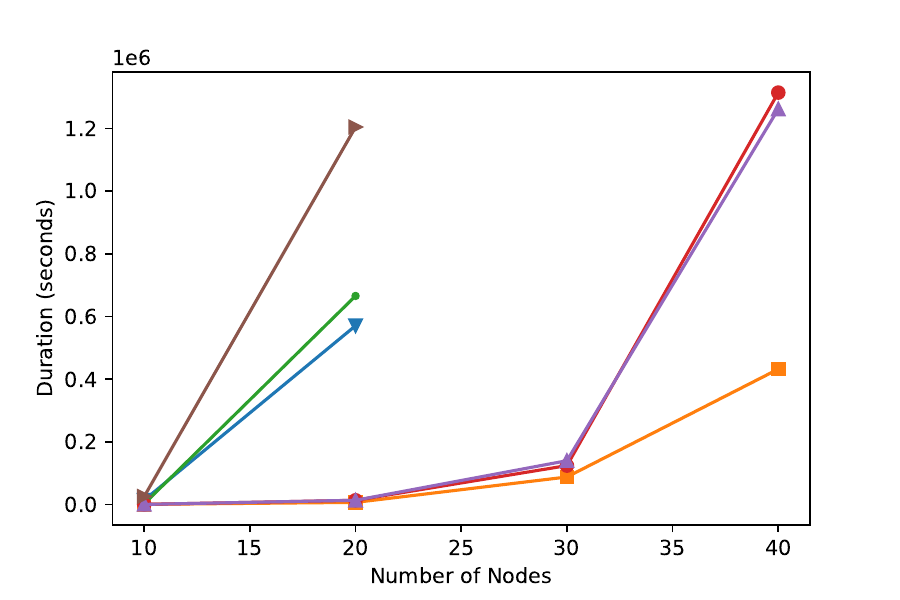}
    \caption{Performance of causal discovery algorithms implemented in \textsc{TETRAD} for semi-synthetic datasets constructed from \cref{alg:semi_synth}. Performance of all methods decreases markedly for larger graphs. Entries are missing if computation time exceeds 36 hours for any method.}\label{fig:benchmarks_TETRAD}
\end{figure}

\begin{sidewaystable}
\centering
\begin{tabular}{| p{2cm}  p{1.5cm} p{0.4cm} p{1.0cm} p{0.4cm} p{0.4cm} |p{0.7cm} p{0.9cm}p{0.7cm} p{0.7cm} p{0.7cm} p{0.7cm} p{0.7cm} p{0.7cm} p{0.9cm} p{2cm}| }
\hline
\multirow{2}{*}[-2pt]{Algorithm}  &\multirow{2}{*}[-2pt]{Data Type}  & \multirow{2}{*}[-2pt]{$p$}  & \multirow{2}{*}[-2pt]{$n$}  & \multirow{2}{*}[-4pt]{$|\mathcal{X}|$}  &  \multirow{2}{*}[-4pt]{$|\mathcal{Y}|$}  &  \multicolumn{10}{|c|}{F1 Score}   \\
& & & & &  &  SCSL & PC-p & PC  & BOSS & CCD & FCI & FGES & 	GFCI & 	GRASP & 	GRaSP-FCI  
 \\
\hline
Real-World & Discrete &  & 200 & 5 & 5 & \bfseries 0.33 & 0.29 & 0.00 & 0.00 & 0.00 & 0.00 & 0.00 & 0.00 & 0.00 & 0.00 \\
 & Discrete &  & 200 & 10 & 10 &\bfseries  0.07 &  0.05 & 0.00 & 0.00 & 0.00 & 0.00 & 0.00 & 0.00 & 0.00 & 0.00 \\
 & Discrete &  & 200 & 15 & 15 & \bfseries 0.09 & 0.07 & 0.00 & 0.00 & 0.00 & 0.00 & 0.03 & 0.03 & 0.03 & 0.06 \\
 & Discrete &  & 200 & 20 & 20 & 0.04 & 0.04 & 0.02 & \bfseries 0.11 & 0.02 & 0.02 & 0.04 & 0.04 & 0.06 & 0.06 \\
& Discrete &  & 2,000 & 5 & 5 & \bfseries 0.71 & 0.48 & 0.00 & 0.18 & 0.00 & 0.00 & 0.17 & 0.17 & 0.00 & 0.17 \\
 & Discrete &  & 2,000 & 10 & 10 & \bfseries 0.28 & 0.15 & 0.00 & 0.00 & 0.00 & 0.00 & 0.00 & 0.00 & 0.00 & 0.00 \\
 & Discrete &  & 2,000 & 15 & 15 & \bfseries 0.13 & 0.11 & 0.00 & 0.03 & 0.00 & 0.00 & 0.00 &  & 0.00 & 0.03 \\
 & Discrete &  & 2,000 & 20 & 20 & 0.04 & \bfseries 0.07 & 0.00 & 0.04 & 0.00 & 0.00 & 0.02 &  & 0.04 & 0.04 \\
 & Discrete &  & 20,000 & 5 & 5 & 0.87 & 0.57 & \bfseries 0.95 & 0.84 & \bfseries 0.95 & 0.82 & \bfseries 0.95 & 0.89 & 0.84 & 0.89 \\
 & Discrete &  & 20,000 & 10 & 10 & \bfseries 0.77 &  & 0.29 & 0.46 & 0.29 & 0.06 & 0.46 & 0.24 & 0.38 & 0.24 \\
& Discrete &  & 20,000 & 15 & 15 & \bfseries 0.44 &  &  & 0.15 &  & 0.00 & 0.12 & 0.00 & 0.15 & 0.06 \\
 & Discrete &  & 20,000 & 20 & 20 & \bfseries 0.31 &  &  & 0.06 &  &  & 0.04 & 0.02 & 0.08 &  \\
 \hline
Synthetic & Discrete & 0.0 & 20,000 & 5 & 5 & \bfseries 0.77 & 0.48 & 0.75 & 0.67 & 0.75 & 0.75 & 0.67 & 0.67 & 0.67 & 0.67 \\
 & Discrete & 0.2 & 20,000 & 5 & 5 & 0.92 & 0.52 & 0.89 & 0.89 & 0.89 & 0.89 & \bfseries 0.95 & 0.89 & 0.89 & 0.89 \\
 & Discrete & 0.4 & 20,000 & 5 & 5 & \bfseries 0.71 & 0.44 & 0.57 & 0.67 & 0.57 & 0.57 & 0.67 & 0.57 & 0.67 & 0.57 \\
& Discrete & 0.6 & 20,000 & 5 & 5 & \bfseries 0.75 & 0.44 & 0.57 & 0.67 & 0.57 & 0.57 & 0.53 & 0.57 & 0.67 & 0.57 \\
 & Discrete & 0.8 & 20,000 & 5 & 5 & \bfseries 0.82 &  & 0.75 & 0.75 & 0.75 & 0.75 & 0.75 & 0.67 & 0.75 & 0.67 \\
 & Discrete & 0.0 & 20,000 & 10 & 10 & \bfseries 0.82 & 0.41 & 0.78 & 0.78 &  & 0.78 & 0.74 &  & 0.77 &  \\
 & Discrete & 0.2 & 20,000 & 10 & 10 & 0.55 & 0.31 & 0.59 & \bfseries 0.60 &  & 0.59 & 0.56 &  & \bfseries 0.60 &  \\
 & Discrete & 0.4 & 20,000 & 10 & 10 & 0.59 & 0.29 & 0.54 & \bfseries 0.70 &  &  & 0.64 &  & \bfseries 0.70 & 0.54 \\
 & Discrete & 0.6 & 20,000 & 10 & 10 & 0.54 & 0.33 & 0.57 & \bfseries 0.67 &  &  & 0.59 &  & \bfseries 0.67 &  \\
 & Discrete & 0.8 & 20,000 & 10 & 10 & 0.57 & 0.34 & 0.54 & \bfseries 0.80 &  &  & 0.57 &  & \bfseries 0.80 &  \\
 & Discrete & 0.0 & 20,000 & 15 & 15 & \bfseries 0.67 &  &  & 0.63 &  &  & 0.63 &  & 0.63 &  \\
 & Discrete & 0.2 & 20,000 & 15 & 15 & 0.68 & 0.32 &  & \bfseries 0.71 &  &  & 0.62 &  & 0.69 &  \\
 & Discrete & 0.4 & 20,000 & 15 & 15 & 0.54 &  &  & \bfseries 0.72 &  &  & 0.48 &  & \bfseries 0.72 &  \\
 & Discrete & 0.6 & 20,000 & 15 & 15 & 0.52 & 0.23 &  & \bfseries 0.78 &  &  & 0.63 &  & 0.76 &  \\
 & Discrete & 0.8 & 20,000 & 15 & 15 & 0.56 & 0.23 &  & \bfseries 0.76 &  &  & 0.51 &  & \bfseries 0.76 &  \\
 & Discrete & 0.0 & 20,000 & 20 & 20 & 0.76 &  &  & \bfseries 0.79 &  &  & 0.77 &  & \bfseries 0.79 &  \\
 & Discrete & 0.2 & 20,000 & 20 & 20 & 0.65 &  &  & 0.82 &  &  & 0.75 &  & \bfseries 0.83 &  \\
 & Discrete & 0.4 & 20,000 & 20 & 20 & 0.42 &  &  & 0.71 &  &  & 0.53 &  & \bfseries 0.72 &  \\
 & Discrete & 0.6 & 20,000 & 20 & 20 & 0.35 &  &  & \bfseries 0.67 &  &  & 0.58 &  & 0.65 &  \\
 & Discrete & 0.8 & 20,000 & 20 & 20 & 0.35 &  &  & \bfseries 0.69 &  &  & 0.63 &  & 0.69 &  \\
\hline
\end{tabular}
\caption{Comparison between F1 scores of SCSL with existing methods for large-scale causal discovery for synthetic datasets with discrete-valued entries of $\mathcal{Y}$. Datasets with real-world confounding tend to favor SCSL, while the results are more mixed when using synthetic confounding.}
\label{tab:discrete_f1}
\end{sidewaystable}

\begin{sidewaystable}
\centering
\begin{tabular}{| p{2cm}  p{1.5cm} p{0.4cm} p{1.0cm} p{0.4cm} p{0.4cm} |p{0.7cm} p{0.9cm}p{0.7cm} p{0.7cm} p{0.7cm} p{0.7cm} p{0.7cm} p{0.7cm} p{0.9cm} p{2cm}| }
\hline
\multirow{2}{*}[-2pt]{Algorithm}  &\multirow{2}{*}[-2pt]{Data Type}  & \multirow{2}{*}[-2pt]{$p$}  & \multirow{2}{*}[-2pt]{$n$}  & \multirow{2}{*}[-4pt]{$|\mathcal{X}|$}  &  \multirow{2}{*}[-4pt]{$|\mathcal{Y}|$}  &  \multicolumn{10}{|c|}{F1 Score}   \\
& & & & &  &  SCSL & PC-p & PC  & BOSS & CCD & FCI & FGES & 	GFCI & 	GRASP & 	GRaSP-FCI  
 \\
\hline
Real-World & Discrete &  & 200 & 5 & 5 & \bfseries 0.33 & 0.29 & 0.00 & 0.00 & 0.00 & 0.00 & 0.00 & 0.00 & 0.00 & 0.00 \\
 & Discrete &  & 200 & 5 & 5 & 0.00 & 0.01 & 0.00 & 0.00 & 0.00 & 0.00 & 0.00 & \bfseries 0.00 & 0.00 & 0.00 \\
 & Discrete &  & 200 & 10 & 10 & 0.00 & 0.04 & 0.00 & 0.00 & 0.00 & 0.00 & 0.00 & \bfseries 0.00 & 0.00 & 0.00 \\
 & Discrete &  & 200 & 15 & 15 & 0.00 & 0.16 & 0.00 & \bfseries 0.00 & 0.00 & 0.00 & 0.00 & 0.00 & 0.00 & 0.00 \\
 & Discrete &  & 200 & 20 & 20 & 0.00 & 0.36 & 0.00 & 0.00 & 0.01 & 0.01 & 0.00 & \bfseries 0.00 & 0.00 & 0.00 \\
 & Discrete &  & 2,000 & 5 & 5 & 0.01 & 0.05 & 0.00 & \bfseries 0.00 & 0.01 & 0.00 & 0.00 & 0.00 & 0.00 & 0.00 \\
 & Discrete &  & 2,000 & 10 & 10 & 0.01 & 0.42 & 0.01 & \bfseries 0.00 & 0.04 & 0.02 & 0.00 & 0.00 & 0.00 & 0.00 \\
 & Discrete &  & 2,000 & 15 & 15 & 0.01 & 1.10 & 0.06 & \bfseries 0.01 & 0.17 & 0.05 & 0.01 &  & 0.01 & 0.01 \\
 & Discrete &  & 2,000 & 20 & 20 & 0.02 & 1.86 & 0.10 & 0.01 & 0.18 & 0.07 & 0.01 &  & 0.01 & \bfseries 0.01 \\
 & Discrete &  & 20,000 & 5 & 5 & 0.16 & 0.73 & 1.44 & 0.01 & 2.86 & 2.94 & \bfseries 0.01 & 1.66 & 0.01 & 1.13 \\
 & Discrete &  & 20,000 & 10 & 10 & 0.23 &  & 9.10 & 0.05 & 13.00 & 8.74 & \bfseries 0.02 & 5.29 & 0.05 & 4.97 \\
 & Discrete &  & 20,000 & 15 & 15 & 0.28 &  &  & 0.11 &  & 22.80 & \bfseries 0.05 & 10.94 & 0.11 & 10.50 \\
 & Discrete &  & 20,000 & 20 & 20 & 0.30 &  &  & 0.21 &  &  & \bfseries 0.11 & 13.86 & 0.19 &  \\
\hline
Synthetic & Discrete & 0.00 & 20,000 & 5 & 5 & 0.07 & 0.19 & 0.13 & 0.01 & 0.53 & 0.22 & \bfseries 0.00 & 0.27 & 0.00 & 0.26 \\
 & Discrete & 0.20 & 20,000 & 5 & 5 & 0.11 & 0.22 & 0.35 & 0.01 & 0.91 & 0.84 & \bfseries 0.00 & 1.08 & 0.01 & 0.55 \\
 & Discrete & 0.40 & 20,000 & 5 & 5 & 0.08 & 0.16 & 0.26 & 0.01 & 1.23 & 0.63 & \bfseries 0.00 & 0.64 & 0.01 & 0.36 \\
 & Discrete & 0.60 & 20,000 & 5 & 5 & 0.09 & 0.19 & 0.30 & 0.01 & 0.97 & 0.55 & 0.01 & 1.36 & \bfseries 0.01 & 0.32 \\
 & Discrete & 0.80 & 20,000 & 5 & 5 & 0.17 &  & 0.21 & 0.01 & 1.47 & 0.46 & 0.01 & 0.70 & \bfseries 0.01 & 0.54 \\
 & Discrete & 0.00 & 20,000 & 10 & 10 & 0.13 & 3.91 & 13.35 & 0.15 &  & 20.22 & \bfseries 0.02 &  & 0.09 &  \\
 & Discrete & 0.20 & 20,000 & 10 & 10 & 0.16 & 3.56 & 8.74 & 0.08 &  & 19.54 & \bfseries 0.02 &  & 0.06 &  \\
 & Discrete & 0.40 & 20,000 & 10 & 10 & 0.22 & 2.87 & 7.30 & 0.12 &  &  & \bfseries 0.03 &  & 0.12 & 18.81 \\
 & Discrete & 0.60 & 20,000 & 10 & 10 & 0.16 & 5.11 & 10.82 & 0.21 &  &  & \bfseries 0.05 &  & 0.22 &  \\
 & Discrete & 0.80 & 20,000 & 10 & 10 & 0.22 & 5.96 & 10.93 & 0.31 &  &  & \bfseries 0.08 &  & 0.28 &  \\
 & Discrete & 0.00 & 20,000 & 15 & 15 & 0.30 &  &  & 0.59 &  &  & \bfseries 0.05 &  & 0.44 &  \\
 & Discrete & 0.20 & 20,000 & 15 & 15 & 0.18 & 16.36 &  & 0.94 &  &  & \bfseries 0.08 &  & 0.66 &  \\
 & Discrete & 0.40 & 20,000 & 15 & 15 & 0.25 &  &  & 1.04 &  &  & \bfseries 0.18 &  & 0.79 &  \\
 & Discrete & 0.60 & 20,000 & 15 & 15 & 0.21 & 17.61 &  & 1.81 &  &  & \bfseries 0.18 &  & 1.50 &  \\
 & Discrete & 0.80 & 20,000 & 15 & 15 & \bfseries 0.26 & 18.47 &  & 2.17 &  &  & 0.27 &  & 1.41 &  \\
 & Discrete & 0.00 & 20,000 & 20 & 20 & 0.33 &  &  & 2.19 &  &  & \bfseries 0.13 &  & 1.67 &  \\
 & Discrete & 0.20 & 20,000 & 20 & 20 & \bfseries 0.23 &  &  & 3.04 &  &  & 0.24 &  & 2.52 &  \\
 & Discrete & 0.40 & 20,000 & 20 & 20 & \bfseries 0.25 &  &  & 6.30 &  &  & 0.57 &  & 3.69 &  \\
 & Discrete & 0.60 & 20,000 & 20 & 20 & \bfseries 0.30 &  &  & 5.51 &  &  & 0.72 &  & 4.26 &  \\
 & Discrete & 0.80 & 20,000 & 20 & 20 & \bfseries 0.24 &  &  & 6.03 &  &  & 0.64 &  & 4.37 &  \\
\hline
\end{tabular}
\caption{Comparison of wall time (in hours) scores of SCSL with existing methods for large-scale causal discovery for  datasets with discrete-valued entries of $\mathcal{Y}$. Datasets with  confounding tend to favor SCSL, while the results are more mixed when using  confounding. Blank entries note that the method did not finish running after 12 hours of computation time.}
\label{tab:discrete_duration}
\end{sidewaystable}

{\color{red}

\begin{sidewaystable}
\centering
\begin{tabular}{| p{2cm}  p{1.5cm} p{0.4cm} p{1.0cm} p{0.4cm} p{0.4cm} |p{0.7cm} p{0.9cm}p{0.7cm} p{0.7cm} p{0.7cm} p{0.7cm} p{0.7cm} p{0.7cm} p{0.9cm} p{2cm}| }
\hline
\multirow{2}{*}[-2pt]{Algorithm}  &\multirow{2}{*}[-2pt]{Data Type}  & \multirow{2}{*}[-2pt]{$p$}  & \multirow{2}{*}[-2pt]{$n$}  & \multirow{2}{*}[-4pt]{$|\mathcal{X}|$}  &  \multirow{2}{*}[-4pt]{$|\mathcal{Y}|$}  &  \multicolumn{10}{|c|}{F1 Score}   \\
& & & & &  &  SCSL & PC-p & PC  & BOSS & CCD & FCI & FGES & 	GFCI & 	GRASP & 	GRaSP-FCI  
 \\
\hline
Synthetic & Continuous & 0.40 & 200 & 5 & 5 & 0.67 & 0.35 & 0.46 & \bfseries 0.84 & 0.46 & 0.46 & \bfseries 0.84 &  & \bfseries 0.84 & 0.57 \\
 & Continuous & 0.40 & 200 & 10 & 10 & 0.24 & 0.24 & 0.45 & 0.75 & 0.45 & 0.45 & \bfseries 0.77 & 0.59 & \bfseries 0.77 & 0.67 \\
 & Continuous & 0.40 & 200 & 15 & 15 & 0.00 & 0.08 & 0.21 & \bfseries 0.75 & 0.21 & 0.21 & 0.45 & 0.23 & 0.75 & 0.39 \\
 & Continuous & 0.40 & 200 & 20 & 20 & 0.10 & 0.06 & 0.25 & \bfseries 0.74 & 0.25 & 0.25 & 0.34 & 0.19 & 0.73 & 0.32 \\
 & Continuous & 0.40 & 2000 & 5 & 5 & 0.89 & 0.55 & 0.82 & \bfseries 1.00 & 0.82 & 0.82 & \bfseries 1.00 &  & \bfseries 1.00 & 0.89 \\
 & Continuous & 0.40 & 2000 & 10 & 10 & 0.00 & 0.27 & 0.24 & \bfseries 0.90 & 0.24 & 0.24 & 0.61 & 0.29 & \bfseries 0.90 & 0.38 \\
 & Continuous & 0.40 & 2000 & 15 & 15 & 0.12 & 0.20 & 0.38 & \bfseries 0.93 & 0.38 & 0.38 & 0.56 & 0.31 & 0.91 & 0.54 \\
 & Continuous & 0.40 & 2000 & 20 & 20 & 0.10 & 0.16 & 0.28 & \bfseries 0.89 & 0.28 & 0.28 & 0.49 &  & 0.89 & 0.43 \\
 & Continuous & 0.00 & 20000 & 5 & 5 & 0.91 & 0.57 & \bfseries 1.00 & \bfseries 1.00 & \bfseries 1.00 & \bfseries 1.00 & \bfseries 1.00 & \bfseries 1.00 & \bfseries 1.00 & \bfseries 1.00 \\
 & Continuous & 0.20 & 20000 & 5 & 5 & \bfseries 1.00 & 0.57 & \bfseries 1.00 & \bfseries 1.00 & \bfseries 1.00 & \bfseries 1.00 & \bfseries 1.00 &  & \bfseries 1.00 & \bfseries 1.00 \\
 & Continuous & 0.40 & 20000 & 5 & 5 & 0.95 & 0.57 & 0.95 & \bfseries 1.00 & 0.95 & 0.95 & 0.95 &  & \bfseries 1.00 & \bfseries 1.00 \\
 & Continuous & 0.60 & 20000 & 5 & 5 & 0.95 & 0.57 & 0.75 & \bfseries 1.00 & 0.75 & 0.75 & \bfseries 1.00 & 0.75 & \bfseries 1.00 & 0.82 \\
& Continuous & 0.80 & 20000 & 5 & 5 & 0.78 & 0.57 & 0.62 & \bfseries 1.00 & 0.62 & 0.67 & 0.53 & 0.62 & \bfseries 1.00 & 0.67 \\
& Continuous & 0.00 & 20000 & 10 & 10 & \bfseries 0.98 & 0.46 & 0.95 & \bfseries 0.98 &  & 0.95 & 0.97 &  &  & 0.97 \\
 & Continuous & 0.20 & 20000 & 10 & 10 & 0.88 & 0.43 & 0.85 & \bfseries 0.98 & 0.85 &  & 0.89 &  & \bfseries 0.98 & 0.91 \\
 & Continuous & 0.40 & 20000 & 10 & 10 & 0.58 &  & 0.75 & \bfseries 1.00 &  & 0.70 & 0.80 &  & \bfseries 1.00 & 0.75 \\
 & Continuous & 0.60 & 20000 & 10 & 10 & 0.08 & 0.35 & 0.48 & \bfseries 0.97 & 0.48 & 0.41 & 0.50 &  & \bfseries 0.97 & 0.60 \\
 & Continuous & 0.80 & 20000 & 10 & 10 & 0.12 & 0.31 & 0.37 & \bfseries 0.98 & 0.37 & 0.32 & 0.48 &  & \bfseries 0.98 & 0.29 \\
 & Continuous & 0.00 & 20000 & 15 & 15 & \bfseries 0.97 &  &  & 0.96 &  &  & 0.90 &  &  &  \\
 & Continuous & 0.20 & 20000 & 15 & 15 & 0.24 &  & 0.55 & 0.95 &  &  & 0.73 &  & \bfseries 0.96 &  \\
 & Continuous & 0.40 & 20000 & 15 & 15 & 0.15 &  & 0.44 & \bfseries 0.98 &  & 0.44 & 0.66 &  & \bfseries 0.98 &  \\
 & Continuous & 0.60 & 20000 & 15 & 15 & 0.12 &  & 0.39 & \bfseries 0.97 &  & 0.39 & 0.55 &  & \bfseries 0.97 &  \\
 & Continuous & 0.80 & 20000 & 15 & 15 & 0.28 &  & 0.32 & \bfseries 0.99 &  &  & 0.57 &  &  &  \\
 & Continuous & 0.00 & 20000 & 20 & 20 & 0.92 &  &  & 0.97 &  &  & 0.92 &  & \bfseries 0.98 &  \\
 & Continuous & 0.20 & 20000 & 20 & 20 & 0.14 &  &  & \bfseries 0.98 &  &  & 0.52 &  & \bfseries 0.98 &  \\
 & Continuous & 0.40 & 20000 & 20 & 20 & 0.17 &  &  & \bfseries 0.98 &  &  & 0.52 &  & \bfseries 0.98 &  \\
 & Continuous & 0.60 & 20000 & 20 & 20 & 0.18 &  &  & \bfseries 1.00 &  &  & 0.41 &  &  &  \\
 & Continuous & 0.80 & 20000 & 20 & 20 & 0.38 &  &  & 0.97 &  &  & 0.38 &  & \bfseries 0.98 &  \\
\hline
\end{tabular}
\caption{Comparison between F1 scores of SCSL with existing methods for large-scale causal discovery for synthetic datasets with continuous-valued entries of $\mathcal{Y}$. SCSL has better performance than other $p$-value producing methods and has comparable performance to FGES. }
\label{tab:continuous_f1}
\end{sidewaystable}

\begin{sidewaystable}
\centering
\begin{tabular}{| p{2cm}  p{1.5cm} p{0.4cm} p{1.0cm} p{0.4cm} p{0.4cm} |p{0.7cm} p{0.9cm}p{0.7cm} p{0.7cm} p{0.7cm} p{0.7cm} p{0.7cm} p{0.7cm} p{0.9cm} p{2cm}| }
\hline
\multirow{2}{*}[-2pt]{Algorithm}  &\multirow{2}{*}[-2pt]{Data Type}  & \multirow{2}{*}[-2pt]{$p$}  & \multirow{2}{*}[-2pt]{$n$}  & \multirow{2}{*}[-4pt]{$|\mathcal{X}|$}  &  \multirow{2}{*}[-4pt]{$|\mathcal{Y}|$}  &  \multicolumn{10}{|c|}{F1 Score}   \\
& & & & &  &  SCSL & PC-p & PC  & BOSS & CCD & FCI & FGES & 	GFCI & 	GRASP & 	GRaSP-FCI  
 \\
\hline
Synthetic & Continuous & 0.40 & 200 & 5 & 5 & 0.00 & 0.05 & 0.00 & 0.00 & 0.00 & 0.00 & 0.00 &  & \bfseries 0.00 & 0.00 \\
 & Continuous & 0.40 & 200 & 10 & 10 & \bfseries 0.00 & 0.22 & 0.01 & 0.00 & 0.01 & 0.00 & 0.00 & 0.00 & 0.00 & 0.01 \\
 & Continuous & 0.40 & 200 & 15 & 15 & \bfseries 0.00 & 0.35 & 0.02 & 0.01 & 0.02 & 0.01 & 0.00 & 0.02 & 0.01 & 0.13 \\
 & Continuous & 0.40 & 200 & 20 & 20 & \bfseries 0.00 & 0.67 & 0.03 & 0.05 & 0.03 & 0.03 & 0.01 & 0.11 & 0.03 & 0.37 \\
 & Continuous & 0.40 & 2,000 & 5 & 5 & 0.01 & 0.17 & 0.01 & 0.00 & 0.02 & 0.01 & 0.00 &  & \bfseries 0.00 & 0.01 \\
& Continuous & 0.40 & 2,000 & 10 & 10 & 0.01 & 1.13 & 0.09 & 0.02 & 0.19 & 0.12 & \bfseries 0.01 & 0.64 & 0.01 & 0.47 \\
 & Continuous & 0.40 & 2,000 & 15 & 15 & \bfseries 0.01 & 3.67 & 0.33 & 0.16 & 0.74 & 0.41 & 0.02 & 14.58 & 0.10 & 0.94 \\
 & Continuous & 0.40 & 2,000 & 20 & 20 & \bfseries 0.01 & 6.30 & 0.70 & 0.69 & 1.37 & 1.02 & 0.09 &  & 0.44 & 11.23 \\
 & Continuous & 0.00 & 20,000 & 5 & 5 & 0.08 & 2.16 & 0.60 & 0.01 & 1.52 & 1.09 & 0.01 & 1.50 & \bfseries 0.01 & 1.72 \\
& Continuous & 0.20 & 20,000 & 5 & 5 & 0.11 & 2.45 & 0.36 & 0.01 & 0.83 & 1.34 & \bfseries 0.01 &  & 0.01 & 0.58 \\
 & Continuous & 0.40 & 20,000 & 5 & 5 & 0.18 & 1.03 & 0.40 & 0.01 & 1.00 & 0.89 & 0.02 &  & \bfseries 0.01 & 0.45 \\
 & Continuous & 0.60 & 20,000 & 5 & 5 & 0.14 & 1.09 & 0.35 & 0.01 & 0.83 & 0.41 & 0.01 & 0.90 & \bfseries 0.01 & 0.35 \\
 & Continuous & 0.80 & 20,000 & 5 & 5 & 0.18 & 1.22 & 0.45 & 0.01 & 1.53 & 0.58 & 0.01 & 3.09 & \bfseries 0.01 & 0.37 \\
 & Continuous & 0.00 & 20,000 & 10 & 10 & 0.18 & 26.54 & 6.00 & 0.14 &  & 25.02 & \bfseries 0.04 &  &  & 22.59 \\
 & Continuous & 0.20 & 20,000 & 10 & 10 & 0.14 & 31.70 & 7.03 & 0.23 & 35.60 &  & \bfseries 0.04 &  & 0.12 & 33.37 \\
 & Continuous & 0.40 & 20,000 & 10 & 10 & 0.24 &  & 4.06 & 0.26 &  & 7.72 & \bfseries 0.05 &  & 0.16 & 14.22 \\
 & Continuous & 0.60 & 20,000 & 10 & 10 & 0.18 & 22.25 & 5.31 & 0.46 & 17.18 & 6.49 & \bfseries 0.12 &  & 0.27 & 12.68 \\
 & Continuous & 0.80 & 20,000 & 10 & 10 & 0.23 & 32.75 & 4.26 & 0.44 & 16.09 & 5.19 & \bfseries 0.10 &  & 0.38 & 14.91 \\
 & Continuous & 0.00 & 20,000 & 15 & 15 & 0.23 &  &  & 0.91 &  &  & \bfseries 0.07 &  &  &  \\
 & Continuous & 0.20 & 20,000 & 15 & 15 & \bfseries 0.17 &  & 29.03 & 1.84 &  &  & 0.27 &  & 1.22 &  \\
 & Continuous & 0.40 & 20,000 & 15 & 15 & \bfseries 0.30 &  & 24.83 & 2.04 &  & 34.56 & 0.34 &  & 1.28 &  \\
 & Continuous & 0.60 & 20,000 & 15 & 15 & \bfseries 0.26 &  & 16.28 & 3.29 &  & 28.29 & 0.41 &  & 2.10 &  \\
 & Continuous & 0.80 & 20,000 & 15 & 15 & \bfseries 0.21 &  & 22.17 & 3.62 &  &  & 0.43 &  &  &  \\
& Continuous & 0.00 & 20,000 & 20 & 20 & 0.22 &  &  & 4.31 &  &  & \bfseries 0.22 &  & 1.98 &  \\
 & Continuous & 0.20 & 20,000 & 20 & 20 & \bfseries 0.30 &  &  & 6.95 &  &  & 1.45 &  & 4.46 &  \\
 & Continuous & 0.40 & 20,000 & 20 & 20 & \bfseries 0.26 &  &  & 9.94 &  &  & 1.55 &  & 5.56 &  \\
 & Continuous & 0.60 & 20,000 & 20 & 20 & \bfseries 0.31 &  &  & 12.58 &  &  & 1.59 &  &  &  \\
 & Continuous & 0.80 & 20,000 & 20 & 20 & \bfseries 0.16 &  &  & 14.07 &  &  & 1.36 &  & 11.42 &  \\

\hline
\end{tabular}
\caption{Comparison of wall time (in hours) scores of SCSL with existing methods for large-scale causal discovery for datasets with continuous-valued entries of $\mathcal{Y}$.}
\label{tab:continuous_duration}
\end{sidewaystable}

}